\documentclass[%
 reprint,
 amsmath,amssymb,
  aps,
prl,superscriptaddress
]{revtex4-1}
\usepackage{xparse}
\usepackage{hyperref}
\usepackage{appendix}
\usepackage{graphicx}% Include figure files
\usepackage{dcolumn}% Align table columns on decimal point
\usepackage{bm}% bold math
\usepackage{xcolor}% coloring text

\newcommand{\veff}{v_\text{eff}}
\usepackage{accents}

\newcommand{\ket}[1]{\ensuremath{\left| #1 \right>}}

\newcommand{\matrixel}[3]{\ensuremath{\left< #1 \vphantom{#3} \right| #2 
\left| #3 \vphantom{#1} \right>}}

\usepackage{bbold}

\newcommand{\beginsupplement}{
\clearpage
\pagebreak
\setcounter{equation}{0}
\setcounter{figure}{0}
\setcounter{table}{0}
\setcounter{page}{1}
\makeatletter
\renewcommand{\theequation}{S\arabic{equation}}
\renewcommand{\thefigure}{S\arabic{figure}}
\onecolumngrid
\section{\large{Supplemental Material}}
}

\hypersetup{
 colorlinks=true,
 linkcolor=blue,
 anchorcolor = blue,
 citecolor = blue,
 filecolor = blue,
 urlcolor = blue
}

\begin{document}

\title{Many-Body Dynamical Localization in a Kicked Lieb-Liniger Gas}
\author{Colin Rylands}
 \email{crylands@umd.edu}
\affiliation{Joint Quantum Institute and
 Condensed Matter Theory Center, University of Maryland, College Park, MD 20742, USA}

 \author{Efim B. Rozenbaum}
 \email{efimroz@umd.edu}
\affiliation{Joint Quantum Institute and
	Condensed Matter Theory Center, University of Maryland, College Park, MD 20742, USA}
\author{Victor Galitski}
\affiliation{Joint Quantum Institute and
	Condensed Matter Theory Center, University of Maryland, College Park, MD 20742, USA}
\author{Robert Konik}
\affiliation{Condensed Matter Physics \& Materials Science Division,
 Brookhaven National Laboratory, Upton, NY 11973-5000, USA}
\date{
    \today
}

\date{\today}
\begin{abstract}
The kicked rotor system is a textbook example of how classical and quantum dynamics can drastically differ. The energy of  a classical particle confined to a ring  and kicked periodically will increase linearly in time whereas in the quantum version the energy saturates after a finite  number of kicks. The quantum system undergoes Anderson localization in angular-momentum space. Conventional wisdom says that in a many-particle system with short-range interactions the localization will be destroyed due to the coupling of widely separated momentum states. Here we provide evidence that for an interacting one-dimensional Bose gas, the Lieb-Liniger model, the dynamical localization can persist.  
\end{abstract}

\maketitle

\textit{Introduction.} \textemdash\, Everyday experience tells us that injecting energy into a closed system causes it to heat up. It follows therefore that doing this repeatedly  will cause the system to heat to infinite temperature. Remarkably this intuition does not necessarily carry over to quantum systems. Recently there has been a large amount of work concerning the prevention of runaway heating in periodically driven closed quantum systems with much of the focus  centered on achieving this via the addition of disorder to the system~\cite{LazaridesDasMoessner, AbaninWojciechHuv,PontePapicAbanin, PonteChandranPapicAbanin}. A far simpler and more intriguing example is provided by the quantum kicked rotor. In this elementary quantum  system a single particle is subjected to a periodic, instantaneous kicking potential, but otherwise propagates freely. After an initial period of increase the energy is seen to saturate, no more energy from the kick can be absorbed, and heating is stopped. This behavior stands in contrast to the corresponding classical system, in which the energy grows without bound,  linearly in time. First discovered numerically~\cite{CasatChiriIzrael,Izrailev80, Chirikov81}, this energy saturation was later elucidated by the construction of a  mapping between the angular-momentum dynamics of the quantum kicked rotor and the dynamics in a lattice model with quasi-disordered potential similar to the Anderson model~\cite{Anderson58}. This mapping shows that the wavefunction becomes exponentially localized in angular-momentum space and leads to the phenomenon being dubbed dynamical localization~\cite{FishmanGrempelPrange, GrempelPrangeFishman}. Subsequently, dynamical localization was observed in clouds of dilute ultra-cold atoms~\cite{Raizen1, Oberthaler, ChabeLemarieDelandaGarreau}.   

A natural question to ask is whether dynamical localization can survive in the presence of interactions. This has been investigated in several studies where interactions have been introduced  through a more complicated kick which couples the particles~\cite{RozenbaumGalitski,Notarnicola} or by including interparticle interactions between the kicks~\cite{Adachi88,Takahashi89, Shepelyansky, PikovskyShepelyansky, Borgonovi95, WenLei09, WenLei10, GligoricBodyfeltFlach, KeserGaneshanRefaelGalitski,QinAndreanovParkFlach,ToikkaAndreanov}. These latter scenarios are of particular interest as interparticle interactions can be readily tuned in ultracold-atom experiments~\cite{BlochDalibardZwerger}.
Using mean-field theory it was shown that after some long time, which is non-linear in the interaction strength, the  kinetic energy of the system grows in a sub-diffusive manner, and localization is destroyed~\cite{Shepelyansky, GligoricBodyfeltFlach}.  Degradation of localization in the presence of interactions has also been shown experimentally in a system of two coupled rotors~\cite{Gadway13}. A lack of heating is also witnessed in other driven interacting quantum systems~ \cite{Dalessio, Bukov,Citro,Rajak}.

In one dimension  perturbative techniques such as mean-field theory break down. Any would-be order, i.e a mean field is destroyed by the strong fluctuations caused by the reduced dimensionality. Systems are strongly correlated as a matter of course, excitations are collective and often cannot be adiabatically connected to the those of free models~\cite{TG, GogolinNerseyanTsvelik}. The description of a kicked interacting Bose gas using mean-field theory is no longer appropriate. Fortunately there exists an array of non-perturbative methods which can be applied to the problem in one dimension.
Here we investigate many-body dynamical localization in an interacting one-dimensional system using a variety of non-perturbative techniques: Fermi-Bose mapping, linear and non-linear Luttinger-Liquid theory, and generalized hydrodynamics~\cite{BertiniColluraDenardisFagotti, CastroDoyon}. We provide evidence that in the presence of interactions one-dimensional systems can dynamically localize. This dynamical localization occurs in the space of many-body eigenstates which results in a saturation of the energy after a finite number of kicks.

%%%%%%%%%%%%%%%%%%%

%%%%%%%%%%%%%%%%%%

\textit{Model.} \textemdash\, 
The system we study consists of an interacting 1D Bose gas which is subjected to a periodic kicking potential. The Hamiltonian which describes this model is a natural extension of the standard single-particle system to the many-body case: 
\begin{eqnarray}\label{H}
H=H_\text{LL}+\sum_{j=-\infty}^\infty\delta(t-jT)H_\text{K}.
\end{eqnarray} 
 The first term is the Lieb-Liniger Hamiltonian~\cite{LiebLiniger1,LiebLiniger2} which provides an excellent description of a 1D cold-atom gas~\cite{Olshanii,DunjkoLorentOlshanii},
\begin{eqnarray}\label{HLL}
\hspace{-7pt}H_\text{LL}\!\!=\!\!\int\mathrm{d}x\,b^\dag(x)\!\!\left[-\frac{\partial_x^2}{2m}\right]\!\!b(x)+c\,b^\dag(x)b(x)b^\dag(x)b(x).
\end{eqnarray}
Here $b^\dag(x)$ and $b(x)$ are creation and annihilation operators, $[b(x),b^\dag(y)]=\delta(x-y)$, describing bosons of mass $m$ which interact with point-like density-density interaction of strength $c\geq 0$ and we have set $\hbar=1$. The model is  integrable and its equilibrium and out-of-equilibrium properties have been extensively studied \cite{ Gaudin, KorepinBook, KormosShashiChouCauxImambekov, DeNardiWoutersBrockmanCaux, CauxKonik}. The eigenstates can be constructed exactly using Bethe Ansatz and are characterised by a set of single-particle momenta, $k_j$, $j=1,\dots \mathcal{N}$ , where $\mathcal{N}$ is the number of particles. The same states are also the eigenstates of an infinite set of non-trivial conserved  operators $Q_n$ ($Q_2\propto H_\text{LL}$) such that $Q_n\ket{\{k_j\}}=\sum_{j=1}^\mathcal{N}k^n_j\ket{\{k_j\}}$. This constrains the dynamics of the system.
The second term in Eq.~\eqref{H} describes the kick which couples to the boson density: 
\begin{eqnarray}
H_\text{K}=\int\mathrm{d}x\,V\cos{(qx)}b^\dag(x)b(x),
\end{eqnarray}
where  $V$ is the kicking strength, $T$ is the kicking period, and $q$ is the wave-vector of the kicking potential. A potential of this form is achieved experimentally by means of a Bragg pulse.

The kicked system follows a two-step time evolution which separates into evolution between the kicks via $e^{-iH_\text{LL}T}$ and  over the kicks via $e^{-iH_\text{K}}$. This can be expressed in terms of a single $H_\text{F}$ known as the Floquet Hamiltonian, governing evolution over one period: $e^{-iH_\text{F}}=e^{-iH_\text{LL}T}e^{-iH_\text{K}}$.
Our goal is to determine the energy of the system after $N$ kicks,
\begin{eqnarray}\label{E}
E(t)=\matrixel{\Psi_0}{e^{iH_\text{F}N}H_{\text{LL}}e^{-iH_\text{F}N}}{\Psi_0},
\end{eqnarray}
 $t=NT$, for some initial state $\ket{\Psi_0}$. Throughout the paper we take the system to be initially in its ground state. 
%  Dynamical localization will be said to occur if the energy, $E(t)$, remains bounded by the some fixed value for arbitrary long time. 

\textit{Tonks-Girardeau limit.} \textemdash\, Aside from the trivial \mbox{$c=0$} limit which recovers the single-particle  model, one can examine the opposite case of  \mbox{$c\to\infty$} known as the the Tonks-Girardeau (TG) gas~\cite{Tonks,Girardeau}. Through Fermi-Bose mapping (FB) the wave-functions of the TG gas take the form of a Slater determinant. This mapping remains valid even in the presence of time-dependent one-body potentials~\cite{YukalovGurardeau, Pezer}. As a result, we may write the solution of the time-dependent Schr{\"o}dinger equation as 
\begin{eqnarray}
\ket{\Psi_0(t)}=\int \mathrm{d}^\mathcal{N}x\,\mathcal{A}\det\left[{\phi_m(x_k,t)}\right]\prod_{l=1}^\mathcal{N}b^\dag(x_l)\ket{0},
\end{eqnarray}
where $\mathcal{A}=\prod_{1\leq i<j\leq\mathcal{N}}\text{sgn}(x_j-x_k)$ is an anti-symmetriser which makes sure the wave-function remains symmetric, and  $\phi_n(x_k,t)$ are a set of orthogonal solutions of the single-particle Schr\"{o}dinger equation $i\partial_t\phi_n(x,t)=\left[-\partial^2_x/2m+\sum\delta(t-jT)V\cos{(qx)}\right]\phi_n(x,t)$. The energy of this state is given by the sum of the single-particle energies,  $E(t)=\sum_{n=1}^\mathcal{N}\int\phi_n^*(x,NT)\left[-\partial^2_x/2m\right]\phi_n(x,NT)$. Since each of the single-particle wave-functions exhibits dynamical localization with the energy remaining bounded, the total energy of the TG gas  will be bounded as well. This proves dynamical localization in the limiting case of a very strongly repulsive Bose gas.

If the system is initially in the ground state all single-particle momentum  states are filled between the Fermi points $|k_j|\leq k_F$, and kicking causes particles to change their momenta by multiples of $q$. Therefore if $q\geq k_F$, particles cannot avoid changing their momenta as a result of the kick. On the other hand if $q=2\pi/L$ then Pauli blocking will  come into play and inhibit the hopping of particles in momentum space. Thus by changing between small and large values of $q$ we can tune between many-body and single-particle physics. Moreover, for any $c\neq 0$, eigenstates of $H_\text{LL}$ obey the Pauli exclusion principle,  i.e. $k_i\neq k_j, \forall i\neq j $~\cite{IzereginKorepin}, so we expect small $q$ to be the most interesting from the perspective of many-body physics.

\textit{Low energy behavior.} \textemdash\, Having established localization at both ends of the range of values for the coupling constant, we turn to a discussion of the system at low energy but for arbitrary $c$. The low-energy behavior of many one-dimensional systems, including the Lieb-Liniger model, is described by the Luttinger-liquid theory~\cite{haldane}. The Hamiltonian of this effective theory can be written in terms of  either bosonic or fermionic fields and for later convenience we choose the latter~\cite{MattisLieb}:
\begin{eqnarray}\nonumber
H_\text{Lutt}&=&\int\mathrm{d}x\,\sum_{\sigma=\pm}:\psi_\sigma^\dag(x)i\sigma v_F\partial_x\psi_\sigma(x):\\\label{HLutt}
&&+g \int\mathrm{d}x:\!\rho_+(x)\!::\!\rho_-(x)\!:\,.
\end{eqnarray}
Here $\psi^\dag_\sigma(x)$ and $\psi_\sigma(x)$ describe right- ($+$) and left- ($-$) moving interacting fermions and $:\ldots :$ denotes normal ordering. 
The fermions also have density-density interactions with strength $g$ which is dependent on $c$. In this language the total density is the sum of left- and right-moving densities, $\rho(x)=\rho_+(x)+\rho_-(x)$, where $\rho_\sigma(x)=\psi_\sigma^\dag(x)\psi_\sigma(x)$, whilst the current is given by the difference $J(x)=\rho_+(x)-\rho_-(x)$.  The kicking term, $H_\text{K}=\int \mathrm{d}x\,V\cos{(qx)} \rho(x)$, therefore separates into terms acting on the left and right movers.

It is possible to bring $H_\text{Lutt}$  to a quadratic form using the unitary transformation $U=e^\Omega$, where
\begin{eqnarray}
\hspace{-10pt}\Omega=\sum_k\frac{\pi \tanh^{-1}(g/2\pi v_F)}{Lk}\left[\tilde{\rho}_{+,-k}\tilde{\rho}_{-,k}-\tilde{\rho}_{-,-k}\tilde{\rho}_{+,k}\right]
\end{eqnarray}
and $\tilde{\rho}_{\sigma,k}$ is the Fourier transform of $\rho_\sigma(x)$~\cite{Rozhkov}. Denoting the transformed operators by a wedge, $\check{\psi}_\sigma=U^\dag\psi_\sigma U$,  we obtain the mapping of the Hamiltonian and the kick to: $H_\text{Lutt}=\int\mathrm{d}x\,\sum_{\sigma=\pm}:\check{\psi}_\sigma^\dag(x)i\sigma v_s\partial_x\check{\psi}_\sigma(x):$ and $H_\text{K}=V\sqrt{K}\int \mathrm{d}x\, \cos{(qx)}\check{\rho}(x)\;$
where $v_s$ is the speed of sound in the system and $K$ is the Luttinger-liquid parameter which depends  on $m$ and $c$ of the original model. In general the relation between $c,m$ and $K,v_s$ must be determined numerically however it is known that for strong repulsive interaction  $K\approx (1+4\rho_0/mc)$ with $\rho_0$ being the average density of the gas, while at weak coupling we have $K\approx\pi \sqrt{\rho_0/mc}$~\cite{Cazalilla}. Thus $K\in[1,\infty]$ whilst $v_s$ is known through the relation $K v_s=v_F$. Note that the effect of the interactions is to modify the kicking strength, $V\to\sqrt{K}V$. The effective kicking strength is larger in the interacting system.

The Luttinger liquid description of the kicked Bose gas relies on the system being initially close to the ground state and  remaining close to it throughout the kicking process $: \Delta E(t)=E(t)-E(0)\ll E_F$. If as a result of the kicking the energy was to increase beyond the purview of the Luttinger-liquid theory, then this would signal a breakdown of our low-energy description, but would not necessarily signal delocalization. We show now that, in fact, for a range of parameters the kicked Luttinger liquid exhibits periodic oscillations of the energy, and the low-energy description remains valid.

By resumming the Baker-Campbell-Hausdorff formula it is possible to determine the Floquet Hamiltonian for the kicked Luttinger liquid exactly. It is given by~\cite{SupplementS2}
\begin{eqnarray}
H_\text{F}=H_\text{Lutt}T+\alpha_\text{K}H_\text{K}+\alpha_JH_J+\kappa,
\end{eqnarray}
with $\alpha_\text{K}=\sin{(v_s q T )}/v_sq T$, $\alpha_{J}=\left[1-\cos{(v_sqT)}\right]/v_sqT$, and $H_J=V\sqrt{K}\int\mathrm{d}x \sin{(qx)}\check{J}(x)$. $\kappa$ is an unimportant constant. 
The Floquet Hamiltonian contains the original unmodified $H_\text{Lutt}$ as well as terms which couple to both the density and current of the system. Eigenstates of $H_\text{F}$ therefore display variations of the density and current on scales $\sim q$. Taking the zero-temperature ground state as $\ket{\Psi_0}$, we find that the change in energy is periodic:
\begin{equation}\label{Luttenergy}
\Delta E(t)=\frac{KV^2L}{v_s\pi T^2}\left[\frac{\sin^2{\left(\frac{v_sq}{2}T\right)}}{v_sqT}\right]\left[\frac{\sin^2{\left(\frac{v_sq}{2}t\right)}}{v_sqT}\right],
\end{equation}
with $L$ being the system size.  Moreover at short time $t\ll 2/v_sq$, it predicts ballistic energy growth $\Delta E(t)\sim t^2$.

This result can be compared with known results for other kicked models which can also be solved exactly \cite{GrempelFishmanPrangeLinear, Berry}.  Therein, quantum-kicked-rotor-like systems with linear dispersion are shown to exhibit bounded dynamics due to integrability of the associated classical model rather than dynamical localization. We emphasize here the distinction between those cases and Eq.~\eqref{Luttenergy}. 
At finite density and low temperature the bare particles of the LL model are completely dissolved by the strong correlations in the system.
The low-energy physics is dictated by collective excitations which can alternately be viewed as sound waves of the Luttinger Liquid or low-momentum quasi-particle-quasi-hole (p-h) excitations near the Fermi surface of the LL model which have linear dispersion $\varepsilon(k)=v_s|k|$. The kicking term creates and destroys p-h excitations only at momenta $\pm q$ 
which  contribute to the periodic oscillation of the energy. Thus in the present case the linear dispersion emerges due to the strongly correlated nature of the system and the self consistency of the approach is guaranteed by the fact that the system is localized.
%%%%%%%%%%%%%%

%%%%%%%%%%%%%

\textit{Non-linear theory}. \textemdash\, To go beyond this low-energy approximation we should include effects of the curvature of the band. This can be readily achieved by working with the fermionic form of the Luttinger liquid~\cite{ImamGlazRMP}. Adding $-\sum_{\sigma}\int\psi_\sigma^\dag(x)\left[\partial_x^2/2m\right]\psi_\sigma(x)$ to Eq.~\eqref{HLutt} and performing the same unitary transformation $U$, we arrive at the following Hamiltonian for the  non-linear Luttinger liquid  ~\cite{Rozhkov}:
\begin{equation}\label{HnLL}
H_\text{nL}\!=\!\sum_{\sigma=\pm}\int\mathrm{d}x:\!\check{\psi}_{\sigma}^\dag(x)\!\left[-i\sigma v_s\partial_x-\frac{\partial^2_x}{2m^*}\right]\!\check{\psi}_\sigma(x)\!:\hspace{1pt}.\hspace{-5pt}
\end{equation}
We see that  the Hamiltonian remains quadratic and the main effect of the interactions is to cause the mass to be renormalised to $1/m^*=v_s/K\partial_\mu(
v_s\sqrt{K})$, where $\mu$ is the chemical potential~\cite{Pereira}. In this description, irrelevant terms which are quartic in the fermions and are higher order in $1/m^*$ have been dropped~\cite{Rozhkov, ImamGlazRMP}. 
We should stress that despite the quadratic nature, Eq.~\eqref{HnLL} contains the effects of the interaction to all orders as well as  the band curvature to the leading order. This approach is the opposite to that of the mean-field theory, where the band curvature is treated exactly and the interactions -- perturbatively. For models with short-range interactions -- like those in the Lieb-Liniger model, Eq.~\eqref{HLL}, -- this method is sufficient to capture the physics beyond the linear regime~\cite{ImamGlazRMP}.
 
The kicking term is unaffected by this new dispersion and so we can describe the gas in terms of  $H_\text{nL}$ at larger values of $V,q$ provided that the kicking does not take the system outside of the regime of a non-linear Luttinger liquid. Once again this is avoided by virtue of the fact that the system dynamically localizes. To see this we note that the full Hamiltonian, including the kick is now no longer integrable as was the case for the linear Luttinger liquid, however it is a quadratic fermionic Hamiltonian, so based on our knowledge of the TG gas we determine that dynamical localization will occur.

%%%%%%%%%%%%%%%%%

%%%%%%%%%%%%%%%%%
\begin{figure}[t]
	\includegraphics[width=\linewidth]{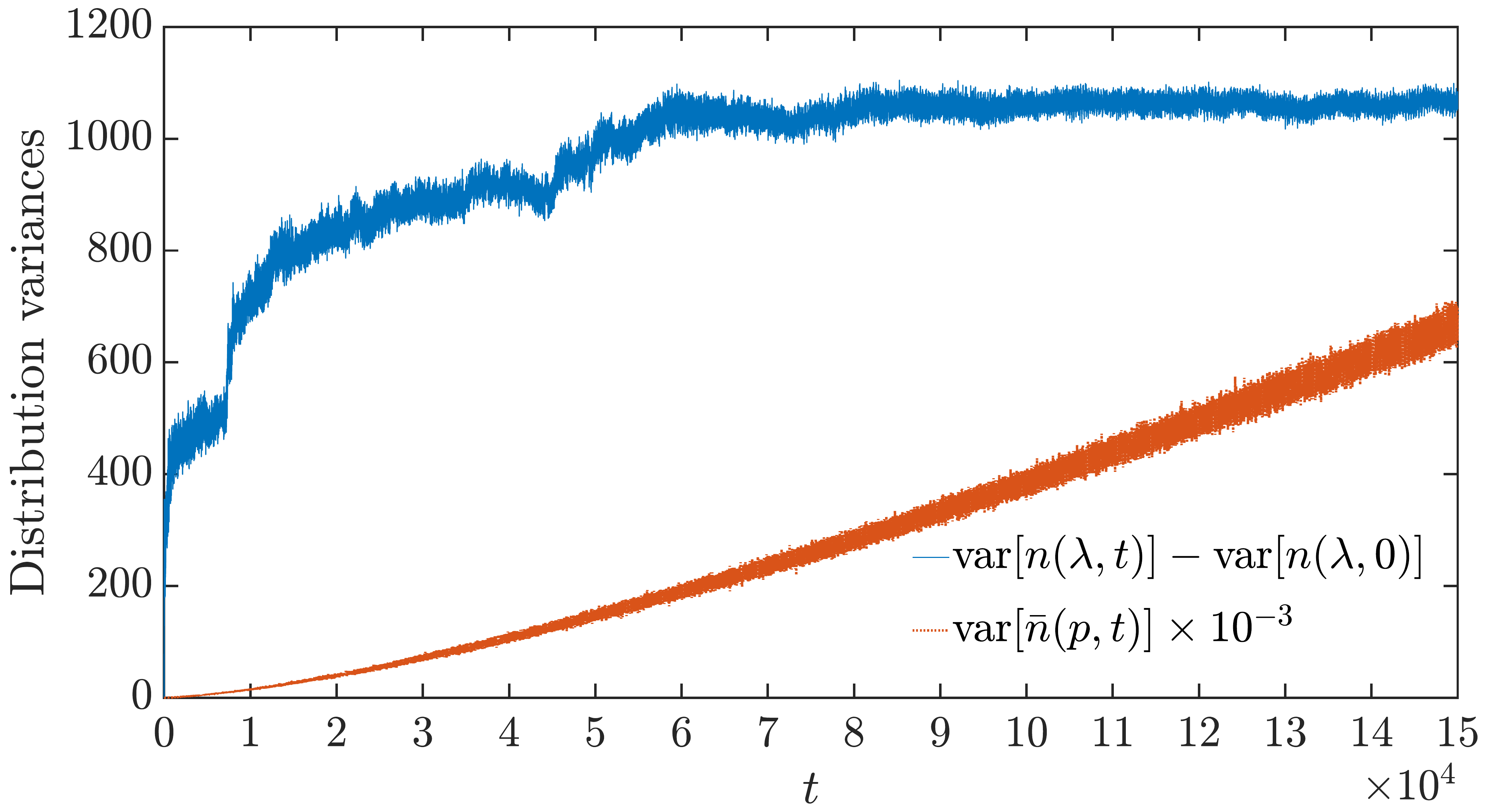}
	\caption{Upper solid blue curve: variance of the momentum density $n(\lambda,t)$ in the kicked Lieb-Liniger gas as a function of time relative to the initial variance: ${\rm var}[n(\lambda,t)]-{\rm var}[n(\lambda,0)]$, where for convenience, we define  ${\rm var}[f(\zeta, t)] = \int \frac{d\zeta}{2\pi}\frac{\zeta^2}{2m}f(\zeta, t)$. It saturates with time, signaling at least transient dynamical localization in sharp contrast to the classical diffusion (heating) under kicking. Lower dotted red curve: scaled variance of $\bar{n}(p,t)$. We were unable to reach its saturation at these parameters, so its continued growth eventually leads to the breakdown of GHD and our scheme and might also signal the potential for eventual delocalization, which we, however, do not observe for a very long time. 
	Parameters: $V=0.5,\; q=4\pi/L,\; \gamma=10,\; \mathcal{N}=200$. { At low enough kicking strength, both variances are well saturated -- see supplemental material for details~\cite{SupplementS4}.}} \label{fig:LL_Energy}
\end{figure}

\textit{Numerical analysis}. \textemdash\,
In order to study the behavior of the system beyond the region of applicability of the analytics, we investigate the   kicked Lieb-Liniger gas numerically, doing so by making use of the integrability of $H_\text{LL}$. The spectrum of the Lieb-Liniger model, as in many other integrable models, consists of long-lived quasi-particles. In the thermodynamic limit and if the variation of the particle density is slow, the system is completely described by the local occupation function of these quasi-particles, $n(x,\lambda,t)$. Here $x$ is the position in space and $\lambda$ is the quasi-particle momentum.  
Generalized hydrodynamics (GHD) is a recently developed theory which describes the evolution of $n(x,\lambda,t)$ at large length scales~\cite{BertiniColluraDenardisFagotti, CastroDoyon}.
Between the kicks the evolution of the gas is determined by the GHD equation:
\begin{eqnarray}\label{BetheBoltz}
\left[\partial_t+\veff\left[n\right]\partial_x\right]n(x,\lambda,t)=0,
\end{eqnarray}
where $\veff[n](x,\lambda,t)$ is the effective velocity of the quasi-particle excitations of the model  which depends upon $n$ itself. With a dressed function $f^\text{dr}(\lambda)$ defined with respect to a bare function $f(\lambda)$ as a solution of  $f^\text{dr}(\lambda)=f(\lambda)+\int\frac{d\mu}{2\pi} \varphi(\lambda-\mu)n(x,\mu,t)f^\text{dr}(\mu)$ with $\varphi(x)=2c/(c^2+x^2)$, the effective velocity is given by 
\begin{equation}\label{eq:veff}
\veff(\lambda)=\left[\varepsilon'(\lambda)\right]^{\text{dr}}/\left[p'(\lambda)\right]^{\text{dr}},
\end{equation}
where $\varepsilon(\lambda)=\lambda^2/2m$ and $p(\lambda)=\lambda$ are the bare energy and momentum of the quasi-particles, and the prime indicates the derivative with respect to $\lambda$.
In both the TG and non-interacting limits, this equation becomes exact~\cite{DoyonDubailKonikYoshimura}, and $n(x,\lambda,t)$ reduces to the Wigner function~\cite{Wigner}. 

To determine the full evolution, we need to compute the effect of the kicks on $n(x,\lambda,t)$. In full generality this is a difficult task which requires the explicit knowledge of the matrix elements of $e^{-iH_\text{K}}$ with arbitrary Lieb-Liniger eigenstates. For slowly varying potential, however, which is an applicability condition of GHD, the situation simplifies. In this case the kicking term couples to the quasi-particles in the same way as to the bare particles described by $b^\dag(x), b(x)$~\cite{DoyonYoshimura}. Hence, over a kick at time $t$ we have~\cite{SupplementS1}:
 \begin{eqnarray}\label{throughkick}
\tilde{n}(x,z,t^+)\!=\!e^{2iV\!\sin{\!(\frac{qz}{2})}\sin{\!(qx)}}\tilde{n}(x,z,t^-),
 \end{eqnarray}
where $\tilde{n}(x,z,t)$ is the Fourier transform of $n(x,\lambda,t)$ with respect to $\lambda$.

\begin{figure}[t]
	\includegraphics[width=\linewidth]{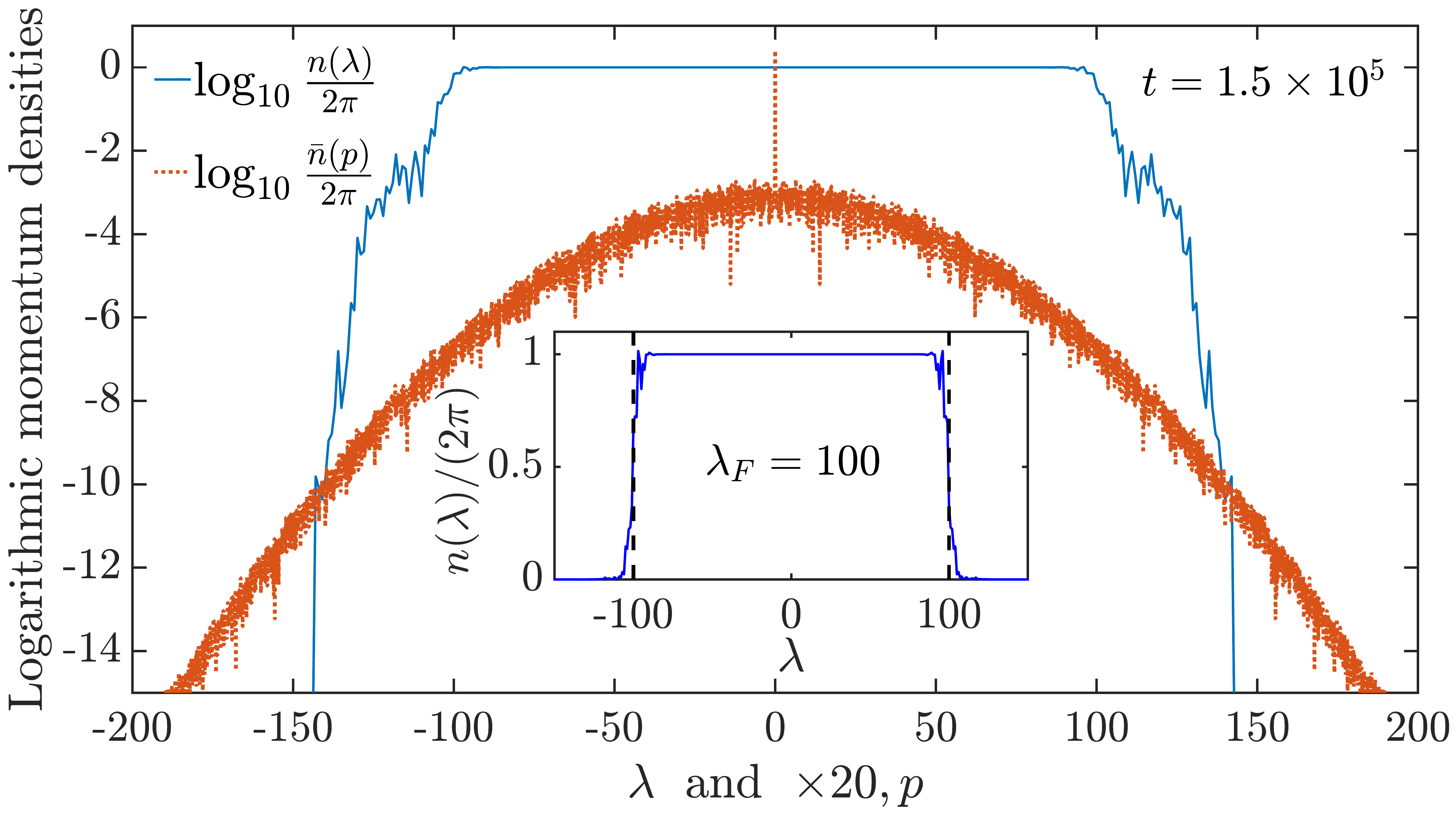}
	\caption{Main plot: decimal logarithms of the momentum densities at the end of the evolution. For $\bar{n}(p)$, all odd momentum components are zero, because we start with the uniform-density initial state and $q=2$ (in units of $2\pi/L$). Only even components of $\bar{n}(p)$ are plotted therefore. Parameters: $V=0.5,\; q=4\pi/L,\; \gamma=10,\; \mathcal{N}=200$. Inset: normalized momentum density $n(\lambda)/(2\pi)$ in the linear scale. The Fermi momentum at our parameter choice is $\lambda_F=100$. %All parameters are the same as in Fig.~\ref{fig:LL_Energy}. Green straight lines show fits of $n(\lambda,\, t=10^5)\propto\exp(\pm\lambda/l_\text{loc})$ with the localization length  $l_\text{loc}\approx 1.67$, which, upon restoring the units, corresponds to $l_\text{loc}\approx 0.27mL/T$. Inset: $n(\lambda,\, t=10^5)$ in the linear scale zoomed in around the remaining Fermi surface. Black dashed lines show Fermi momentum $\lambda_F=100$.
	} \label{fig:LL_MomDens}
\end{figure}
Using Eqs.~\eqref{BetheBoltz} and \eqref{throughkick} we can determine  the total quasi-particle occupation  function, $n(\lambda,t)=\hspace{-1pt}\int\hspace{-1pt}\mathrm{d}x\, n(x,\lambda,t)$, and the energy  $E(t)=\int\frac{\mathrm{d}\lambda}{2\pi}\,\varepsilon^\text{dr}(\lambda)n(\lambda,t)$ of the gas. { We also introduce a common measure of localization, the variance of the occupation function, var$ [n(\lambda,t]$ where ${\rm var}[f(\zeta, t)] = \int \frac{d\zeta}{2\pi}\frac{\zeta^2}{2m}f(\zeta, t)$ (see, e.g., Refs.~\cite{GrempelFishmanPrangeLinear, *Berry, Fishman84, *Prange84, *Lima91, *Borgonovi97}). Saturation of the variance indicates exponential localization in $\lambda$ space.} All the quantities in our calculations are dimensionless and sometimes implicitly expressed in units of $m$, $L/2\pi$, and $T$. 
The evolution between the kicks can be evaluated via a finite-difference scheme $n(x,\lambda,t+\delta t)=n(x- v_\text{eff}[n(x,\lambda,t)]\delta t,\lambda,t)$~\cite{BulchandaniVasseur}, where we choose $T/\delta t=1000$. At each time step $v_\text{eff}$ is reevaluated via Eq.~(\ref{eq:veff}), and the shift is performed in the Fourier space by explicitly calculating the integral $n(x,\lambda,t+\delta t)=
\int dp e^{ip\left\{x-v_\text{eff}[n(x, \lambda, t)]\delta t \right\}}\bar{n}(p,\lambda,t)$, where $\bar{n}(p,\lambda,t)$ is the Fourier transform of $n(x,\lambda,t)$ with respect to $x$. This scheme works well at short times, but due to its very high numerical complexity, for practical purposes we employ a different approach -- a linearized approximation to Eq.~\eqref{BetheBoltz}~\cite{PanfilPawelczyk}. In this approximation we calculate $v_\text{eff}[\left<n\right>]$ after a kick using a spatially averaged $\left<n(\lambda,t)\right>=\int\mathrm{d}x n(x,\lambda,t)/L$ which is then used to propagate the solution over an entire duration of the free evolution at once. This is easily carried out in Fourier space via: $\bar{n}(p,\lambda,t+T)=e^{-ipv_\text{eff}[\left< n\right>]T}\bar{n}(p,\lambda,t)$. The next kick is then applied via Eq.~(\ref{throughkick}), $v_\text{eff}[\left<n\right>]$ is determined anew, and the process is repeated. This approximation becomes exact in the TG case and agrees well with the finite difference scheme at short times. {We demonstrate the validity of this approximation in the supplemental material~\cite{SupplementS4}.}

We choose a small value of the kicking-potential wave-vector $q = 4\pi/L$ and take $V=0.5$. In this case, the corresponding single-particle classical system is in the mixed regular-chaotic regime with the unbounded chaotic sea. The critical value of the kicking strength where the regular-to-chaotic transition occurs is $(qL/2\pi)^2V_\text{cr}\approx0.97$ with the remaining KAM islands vanishing to the naked eye towards $(qL/2\pi)^2V=5$~\cite{Chirikov79,  SupplementS3}. As was seen in the Luttinger liquid analysis, interactions will cause the effective kicking strength to be larger and the analogous critical value to be lower.
 
Fig.~\ref{fig:LL_Energy} shows the momentum variance of the Lieb-Liniger gas under kicking for $\gamma=\mathcal{N}/(mcL)=10$. At short times, the energy grows quickly, but later, it saturates and becomes bounded due to dynamical localization. At the same time,  $n(\lambda,t)$, which is initially the Fermi-Dirac $\Pi$-shaped function with the Fermi momentum $\lambda_F=100$ -- with our choice of $\mathcal{N}=200$, -- acquires exponential tails (see Fig.~\ref{fig:LL_MomDens}) and stops spreading any further after the saturation of energy is reached.  
Fig.~\ref{fig:LL_MomDens} also shows the Fourier transform of the spatial density  $\bar{n}(p,\,t=1.5\times10^5) = \int\frac{\mathrm{d}\lambda}{2\pi}\, \bar{n}(p,\lambda,t=1.5\times10^5)$ that decays exponentially, as well, but its width keeps growing with time, as opposed to the width of $n(\lambda,t)$ -- see Fig.~\ref{fig:LL_Energy}. We were unable to reach its saturation at these parameters, so its continued growth eventually leads to a breakdown of the numerical method and the applicability of GHD. Prior to  this, however, no delocalization is seen for a very long time. { At low enough kicking strength, however, both variances are well saturated. We show that behavior at the kicking strength $V=0.15$ in the supplemental material~\cite{SupplementS4}.}

Before concluding  we wish to emphasize that our results show that a kicked interacting 1D bose gas can exhibit dynamical localisation over certain timescales and provided the system is initiated close to its ground state. Such conditions can be met within cold atom gas systems. This however does not rule out the possibility of delocalization at longer time scales or beyond the applicability of our methods e.g. high temperature or larger kicking strength.
\acknowledgements
The authors are grateful to Jean-Claude Garreau for fruitful discussions. This work was supported by NSF DMR-1613029, US-ARO (contract No. W911NF1310172), DARPA DRINQS program (C.R. and E.R.), DOE-BES (DESC0001911), and the Simons Foundation (V.G.). R.M.K. was supported by the U.S. Department of Energy, Office of Basic Energy Sciences, under Contract No. DE-AC02-98CH10886. The authors acknowledge the University of Maryland supercomputing resources (http://hpcc.umd.edu) made available for conducting the research reported in this paper. 

\bibliography{bib}

\begin{thebibliography}{72}%
\makeatletter
\providecommand \@ifxundefined [1]{%
 \@ifx{#1\undefined}
}%
\providecommand \@ifnum [1]{%
 \ifnum #1\expandafter \@firstoftwo
 \else \expandafter \@secondoftwo
 \fi
}%
\providecommand \@ifx [1]{%
 \ifx #1\expandafter \@firstoftwo
 \else \expandafter \@secondoftwo
 \fi
}%
\providecommand \natexlab [1]{#1}%
\providecommand \enquote  [1]{``#1''}%
\providecommand \bibnamefont  [1]{#1}%
\providecommand \bibfnamefont [1]{#1}%
\providecommand \citenamefont [1]{#1}%
\providecommand \href@noop [0]{\@secondoftwo}%
\providecommand \href [0]{\begingroup \@sanitize@url \@href}%
\providecommand \@href[1]{\@@startlink{#1}\@@href}%
\providecommand \@@href[1]{\endgroup#1\@@endlink}%
\providecommand \@sanitize@url [0]{\catcode `\\12\catcode `\$12\catcode
  `\&12\catcode `\#12\catcode `\^12\catcode `\_12\catcode `\%12\relax}%
\providecommand \@@startlink[1]{}%
\providecommand \@@endlink[0]{}%
\providecommand \url  [0]{\begingroup\@sanitize@url \@url }%
\providecommand \@url [1]{\endgroup\@href {#1}{\urlprefix }}%
\providecommand \urlprefix  [0]{URL }%
\providecommand \Eprint [0]{\href }%
\providecommand \doibase [0]{http://dx.doi.org/}%
\providecommand \selectlanguage [0]{\@gobble}%
\providecommand \bibinfo  [0]{\@secondoftwo}%
\providecommand \bibfield  [0]{\@secondoftwo}%
\providecommand \translation [1]{[#1]}%
\providecommand \BibitemOpen [0]{}%
\providecommand \bibitemStop [0]{}%
\providecommand \bibitemNoStop [0]{.\EOS\space}%
\providecommand \EOS [0]{\spacefactor3000\relax}%
\providecommand \BibitemShut  [1]{\csname bibitem#1\endcsname}%
\let\auto@bib@innerbib\@empty
%</preamble>
\bibitem [{\citenamefont {Lazarides}\ \emph {et~al.}(2014)\citenamefont
  {Lazarides}, \citenamefont {Das},\ and\ \citenamefont
  {Moessner}}]{LazaridesDasMoessner}%
  \BibitemOpen
  \bibfield  {author} {\bibinfo {author} {\bibfnamefont {A.}~\bibnamefont
  {Lazarides}}, \bibinfo {author} {\bibfnamefont {A.}~\bibnamefont {Das}}, \
  and\ \bibinfo {author} {\bibfnamefont {R.}~\bibnamefont {Moessner}},\ }\href
  {\doibase 10.1103/PhysRevLett.112.150401} {\bibfield  {journal} {\bibinfo
  {journal} {Phys. Rev. Lett.}\ }\textbf {\bibinfo {volume} {112}},\ \bibinfo
  {pages} {150401} (\bibinfo {year} {2014})}\BibitemShut {NoStop}%
\bibitem [{\citenamefont {Abanin}\ \emph {et~al.}(2016)\citenamefont {Abanin},
  \citenamefont {Roeck},\ and\ \citenamefont {Huveneers}}]{AbaninWojciechHuv}%
  \BibitemOpen
  \bibfield  {author} {\bibinfo {author} {\bibfnamefont {D.~A.}\ \bibnamefont
  {Abanin}}, \bibinfo {author} {\bibfnamefont {W.~D.}\ \bibnamefont {Roeck}}, \
  and\ \bibinfo {author} {\bibfnamefont {F.}~\bibnamefont {Huveneers}},\ }\href
  {\doibase https://doi.org/10.1016/j.aop.2016.03.010} {\bibfield  {journal}
  {\bibinfo  {journal} {Annals of Physics}\ }\textbf {\bibinfo {volume}
  {372}},\ \bibinfo {pages} {1 } (\bibinfo {year} {2016})}\BibitemShut
  {NoStop}%
\bibitem [{\citenamefont {Ponte}\ \emph
  {et~al.}(2015{\natexlab{a}})\citenamefont {Ponte}, \citenamefont {Papic},
  \citenamefont {Huveneers},\ and\ \citenamefont {Abanin}}]{PontePapicAbanin}%
  \BibitemOpen
  \bibfield  {author} {\bibinfo {author} {\bibfnamefont {P.}~\bibnamefont
  {Ponte}}, \bibinfo {author} {\bibfnamefont {Z.}~\bibnamefont {Papic}},
  \bibinfo {author} {\bibfnamefont {F.}~\bibnamefont {Huveneers}}, \ and\
  \bibinfo {author} {\bibfnamefont {D.~A.}\ \bibnamefont {Abanin}},\ }\href
  {\doibase 10.1103/PhysRevLett.114.140401} {\bibfield  {journal} {\bibinfo
  {journal} {Phys. Rev. Lett.}\ }\textbf {\bibinfo {volume} {114}},\ \bibinfo
  {pages} {140401} (\bibinfo {year} {2015}{\natexlab{a}})}\BibitemShut
  {NoStop}%
\bibitem [{\citenamefont {Ponte}\ \emph
  {et~al.}(2015{\natexlab{b}})\citenamefont {Ponte}, \citenamefont {Chandran},
  \citenamefont {Papi\'{c}},\ and\ \citenamefont
  {Abanin}}]{PonteChandranPapicAbanin}%
  \BibitemOpen
  \bibfield  {author} {\bibinfo {author} {\bibfnamefont {P.}~\bibnamefont
  {Ponte}}, \bibinfo {author} {\bibfnamefont {A.}~\bibnamefont {Chandran}},
  \bibinfo {author} {\bibfnamefont {Z.}~\bibnamefont {Papi\'{c}}}, \ and\
  \bibinfo {author} {\bibfnamefont {D.~A.}\ \bibnamefont {Abanin}},\ }\href
  {\doibase 10.1016/j.aop.2014.11.008} {\bibfield  {journal} {\bibinfo
  {journal} {Ann. Phys.}\ }\textbf {\bibinfo {volume} {353}},\ \bibinfo {pages}
  {196 } (\bibinfo {year} {2015}{\natexlab{b}})}\BibitemShut {NoStop}%
\bibitem [{\citenamefont {{Casati}}\ \emph {et~al.}(1979)\citenamefont
  {{Casati}}, \citenamefont {{Chirikov}}, \citenamefont {{Izraelev}},\ and\
  \citenamefont {{Ford}}}]{CasatChiriIzrael}%
  \BibitemOpen
  \bibfield  {author} {\bibinfo {author} {\bibfnamefont {G.}~\bibnamefont
  {{Casati}}}, \bibinfo {author} {\bibfnamefont {B.~V.}\ \bibnamefont
  {{Chirikov}}}, \bibinfo {author} {\bibfnamefont {F.~M.}\ \bibnamefont
  {{Izraelev}}}, \ and\ \bibinfo {author} {\bibfnamefont {J.}~\bibnamefont
  {{Ford}}},\ }in\ \href {\doibase 10.1007/BFb0021757} {\emph {\bibinfo
  {booktitle} {Stochastic Behavior in Classical and Quantum Hamiltonian
  Systems}}},\ \bibinfo {series} {Lecture Notes in Physics, Berlin Springer
  Verlag}, Vol.~\bibinfo {volume} {93},\ \bibinfo {editor} {edited by\ \bibinfo
  {editor} {\bibfnamefont {G.}~\bibnamefont {{Casati}}}\ and\ \bibinfo {editor}
  {\bibfnamefont {J.}~\bibnamefont {{Ford}}}}\ (\bibinfo {year} {1979})\ pp.\
  \bibinfo {pages} {334--352}\BibitemShut {NoStop}%
\bibitem [{\citenamefont {Izrailev}\ and\ \citenamefont
  {Shepelyansky}(1980)}]{Izrailev80}%
  \BibitemOpen
  \bibfield  {author} {\bibinfo {author} {\bibfnamefont {F.~M.}\ \bibnamefont
  {Izrailev}}\ and\ \bibinfo {author} {\bibfnamefont {D.~L.}\ \bibnamefont
  {Shepelyansky}},\ }\href {\doibase 10.1007/BF01029131} {\bibfield  {journal}
  {\bibinfo  {journal} {Theor. Math. Phys.}\ }\textbf {\bibinfo {volume}
  {43}},\ \bibinfo {pages} {553} (\bibinfo {year} {1980})},\ \bibinfo {note}
  {[Teor. Mat. Fiz. { 43}, 417 (1980)]}\BibitemShut {NoStop}%
\bibitem [{\citenamefont {Chirikov}\ \emph {et~al.}(1981)\citenamefont
  {Chirikov}, \citenamefont {Izrailev},\ and\ \citenamefont
  {Shepelyansky}}]{Chirikov81}%
  \BibitemOpen
  \bibfield  {author} {\bibinfo {author} {\bibfnamefont {B.~V.}\ \bibnamefont
  {Chirikov}}, \bibinfo {author} {\bibfnamefont {F.~M.}\ \bibnamefont
  {Izrailev}}, \ and\ \bibinfo {author} {\bibfnamefont {D.~L.}\ \bibnamefont
  {Shepelyansky}},\ }\href@noop {} {\bibfield  {journal} {\bibinfo  {journal}
  {Sov. Sci. Rev.}\ }\textbf {\bibinfo {volume} {Sec. C 2}},\ \bibinfo {pages}
  {209} (\bibinfo {year} {1981})}\BibitemShut {NoStop}%
\bibitem [{\citenamefont {Anderson}(1958)}]{Anderson58}%
  \BibitemOpen
  \bibfield  {author} {\bibinfo {author} {\bibfnamefont {P.~W.}\ \bibnamefont
  {Anderson}},\ }\href {\doibase 10.1103/PhysRev.109.1492} {\bibfield
  {journal} {\bibinfo  {journal} {Phys. Rev.}\ }\textbf {\bibinfo {volume}
  {109}},\ \bibinfo {pages} {1492} (\bibinfo {year} {1958})}\BibitemShut
  {NoStop}%
\bibitem [{\citenamefont {Fishman}\ \emph {et~al.}(1982)\citenamefont
  {Fishman}, \citenamefont {Grempel},\ and\ \citenamefont
  {Prange}}]{FishmanGrempelPrange}%
  \BibitemOpen
  \bibfield  {author} {\bibinfo {author} {\bibfnamefont {S.}~\bibnamefont
  {Fishman}}, \bibinfo {author} {\bibfnamefont {D.~R.}\ \bibnamefont
  {Grempel}}, \ and\ \bibinfo {author} {\bibfnamefont {R.~E.}\ \bibnamefont
  {Prange}},\ }\href {\doibase 10.1103/PhysRevLett.49.509} {\bibfield
  {journal} {\bibinfo  {journal} {Phys. Rev. Lett.}\ }\textbf {\bibinfo
  {volume} {49}},\ \bibinfo {pages} {509} (\bibinfo {year} {1982})}\BibitemShut
  {NoStop}%
\bibitem [{\citenamefont {Grempel}\ \emph {et~al.}(1984)\citenamefont
  {Grempel}, \citenamefont {Prange},\ and\ \citenamefont
  {Fishman}}]{GrempelPrangeFishman}%
  \BibitemOpen
  \bibfield  {author} {\bibinfo {author} {\bibfnamefont {D.~R.}\ \bibnamefont
  {Grempel}}, \bibinfo {author} {\bibfnamefont {R.~E.}\ \bibnamefont {Prange}},
  \ and\ \bibinfo {author} {\bibfnamefont {S.}~\bibnamefont {Fishman}},\ }\href
  {\doibase 10.1103/PhysRevA.29.1639} {\bibfield  {journal} {\bibinfo
  {journal} {Phys. Rev. A}\ }\textbf {\bibinfo {volume} {29}},\ \bibinfo
  {pages} {1639} (\bibinfo {year} {1984})}\BibitemShut {NoStop}%
\bibitem [{\citenamefont {Moore}\ \emph {et~al.}(1995)\citenamefont {Moore},
  \citenamefont {Robinson}, \citenamefont {Bharucha}, \citenamefont
  {Sundaram},\ and\ \citenamefont {Raizen}}]{Raizen1}%
  \BibitemOpen
  \bibfield  {author} {\bibinfo {author} {\bibfnamefont {F.~L.}\ \bibnamefont
  {Moore}}, \bibinfo {author} {\bibfnamefont {J.~C.}\ \bibnamefont {Robinson}},
  \bibinfo {author} {\bibfnamefont {C.~F.}\ \bibnamefont {Bharucha}}, \bibinfo
  {author} {\bibfnamefont {B.}~\bibnamefont {Sundaram}}, \ and\ \bibinfo
  {author} {\bibfnamefont {M.~G.}\ \bibnamefont {Raizen}},\ }\href {\doibase
  10.1103/PhysRevLett.75.4598} {\bibfield  {journal} {\bibinfo  {journal}
  {Phys. Rev. Lett.}\ }\textbf {\bibinfo {volume} {75}},\ \bibinfo {pages}
  {4598} (\bibinfo {year} {1995})}\BibitemShut {NoStop}%
\bibitem [{\citenamefont {d'Arcy}\ \emph {et~al.}(2001)\citenamefont {d'Arcy},
  \citenamefont {Godun}, \citenamefont {Oberthaler}, \citenamefont
  {Cassettari},\ and\ \citenamefont {Summy}}]{Oberthaler}%
  \BibitemOpen
  \bibfield  {author} {\bibinfo {author} {\bibfnamefont {M.~B.}\ \bibnamefont
  {d'Arcy}}, \bibinfo {author} {\bibfnamefont {R.~M.}\ \bibnamefont {Godun}},
  \bibinfo {author} {\bibfnamefont {M.~K.}\ \bibnamefont {Oberthaler}},
  \bibinfo {author} {\bibfnamefont {D.}~\bibnamefont {Cassettari}}, \ and\
  \bibinfo {author} {\bibfnamefont {G.~S.}\ \bibnamefont {Summy}},\ }\href
  {\doibase 10.1103/PhysRevLett.87.074102} {\bibfield  {journal} {\bibinfo
  {journal} {Phys. Rev. Lett.}\ }\textbf {\bibinfo {volume} {87}},\ \bibinfo
  {pages} {074102} (\bibinfo {year} {2001})}\BibitemShut {NoStop}%
\bibitem [{\citenamefont {Chab\'e}\ \emph {et~al.}(2008)\citenamefont
  {Chab\'e}, \citenamefont {Lemari\'e}, \citenamefont {Gr\'emaud},
  \citenamefont {Delande}, \citenamefont {Szriftgiser},\ and\ \citenamefont
  {Garreau}}]{ChabeLemarieDelandaGarreau}%
  \BibitemOpen
  \bibfield  {author} {\bibinfo {author} {\bibfnamefont {J.}~\bibnamefont
  {Chab\'e}}, \bibinfo {author} {\bibfnamefont {G.}~\bibnamefont {Lemari\'e}},
  \bibinfo {author} {\bibfnamefont {B.}~\bibnamefont {Gr\'emaud}}, \bibinfo
  {author} {\bibfnamefont {D.}~\bibnamefont {Delande}}, \bibinfo {author}
  {\bibfnamefont {P.}~\bibnamefont {Szriftgiser}}, \ and\ \bibinfo {author}
  {\bibfnamefont {J.~C.}\ \bibnamefont {Garreau}},\ }\href {\doibase
  10.1103/PhysRevLett.101.255702} {\bibfield  {journal} {\bibinfo  {journal}
  {Phys. Rev. Lett.}\ }\textbf {\bibinfo {volume} {101}},\ \bibinfo {pages}
  {255702} (\bibinfo {year} {2008})}\BibitemShut {NoStop}%
\bibitem [{\citenamefont {Rozenbaum}\ and\ \citenamefont
  {Galitski}(2017)}]{RozenbaumGalitski}%
  \BibitemOpen
  \bibfield  {author} {\bibinfo {author} {\bibfnamefont {E.~B.}\ \bibnamefont
  {Rozenbaum}}\ and\ \bibinfo {author} {\bibfnamefont {V.}~\bibnamefont
  {Galitski}},\ }\href {\doibase 10.1103/PhysRevB.95.064303} {\bibfield
  {journal} {\bibinfo  {journal} {Phys. Rev. B}\ }\textbf {\bibinfo {volume}
  {95}},\ \bibinfo {pages} {064303} (\bibinfo {year} {2017})}\BibitemShut
  {NoStop}%
\bibitem [{\citenamefont {Notarnicola}\ \emph {et~al.}(2018)\citenamefont
  {Notarnicola}, \citenamefont {Iemini}, \citenamefont {Rossini}, \citenamefont
  {Fazio}, \citenamefont {Silva},\ and\ \citenamefont
  {Russomanno}}]{Notarnicola}%
  \BibitemOpen
  \bibfield  {author} {\bibinfo {author} {\bibfnamefont {S.}~\bibnamefont
  {Notarnicola}}, \bibinfo {author} {\bibfnamefont {F.}~\bibnamefont {Iemini}},
  \bibinfo {author} {\bibfnamefont {D.}~\bibnamefont {Rossini}}, \bibinfo
  {author} {\bibfnamefont {R.}~\bibnamefont {Fazio}}, \bibinfo {author}
  {\bibfnamefont {A.}~\bibnamefont {Silva}}, \ and\ \bibinfo {author}
  {\bibfnamefont {A.}~\bibnamefont {Russomanno}},\ }\href {\doibase
  10.1103/PhysRevE.97.022202} {\bibfield  {journal} {\bibinfo  {journal} {Phys.
  Rev. E}\ }\textbf {\bibinfo {volume} {97}},\ \bibinfo {pages} {022202}
  (\bibinfo {year} {2018})}\BibitemShut {NoStop}%
\bibitem [{\citenamefont {Adachi}\ \emph {et~al.}(1988)\citenamefont {Adachi},
  \citenamefont {Toda},\ and\ \citenamefont {Ikeda}}]{Adachi88}%
  \BibitemOpen
  \bibfield  {author} {\bibinfo {author} {\bibfnamefont {S.}~\bibnamefont
  {Adachi}}, \bibinfo {author} {\bibfnamefont {M.}~\bibnamefont {Toda}}, \ and\
  \bibinfo {author} {\bibfnamefont {K.}~\bibnamefont {Ikeda}},\ }\href
  {\doibase 10.1103/PhysRevLett.61.659} {\bibfield  {journal} {\bibinfo
  {journal} {Phys. Rev. Lett.}\ }\textbf {\bibinfo {volume} {61}},\ \bibinfo
  {pages} {659} (\bibinfo {year} {1988})}\BibitemShut {NoStop}%
\bibitem [{\citenamefont {Takahashi}(1989)}]{Takahashi89}%
  \BibitemOpen
  \bibfield  {author} {\bibinfo {author} {\bibfnamefont {K.}~\bibnamefont
  {Takahashi}},\ }\href {\doibase 10.1143/PTPS.98.109} {\bibfield  {journal}
  {\bibinfo  {journal} {Prog. Theor. Phys. Suppl.}\ }\textbf {\bibinfo {volume}
  {98}},\ \bibinfo {pages} {109} (\bibinfo {year} {1989})}\BibitemShut
  {NoStop}%
\bibitem [{\citenamefont {Shepelyansky}(1993)}]{Shepelyansky}%
  \BibitemOpen
  \bibfield  {author} {\bibinfo {author} {\bibfnamefont {D.~L.}\ \bibnamefont
  {Shepelyansky}},\ }\href {\doibase 10.1103/PhysRevLett.70.1787} {\bibfield
  {journal} {\bibinfo  {journal} {Phys. Rev. Lett.}\ }\textbf {\bibinfo
  {volume} {70}},\ \bibinfo {pages} {1787} (\bibinfo {year}
  {1993})}\BibitemShut {NoStop}%
\bibitem [{\citenamefont {Pikovsky}\ and\ \citenamefont
  {Shepelyansky}(2008)}]{PikovskyShepelyansky}%
  \BibitemOpen
  \bibfield  {author} {\bibinfo {author} {\bibfnamefont {A.~S.}\ \bibnamefont
  {Pikovsky}}\ and\ \bibinfo {author} {\bibfnamefont {D.~L.}\ \bibnamefont
  {Shepelyansky}},\ }\href {\doibase 10.1103/PhysRevLett.100.094101} {\bibfield
   {journal} {\bibinfo  {journal} {Phys. Rev. Lett.}\ }\textbf {\bibinfo
  {volume} {100}},\ \bibinfo {pages} {094101} (\bibinfo {year}
  {2008})}\BibitemShut {NoStop}%
\bibitem [{\citenamefont {Borgonovi}\ and\ \citenamefont
  {Shepelyansky}(1995)}]{Borgonovi95}%
  \BibitemOpen
  \bibfield  {author} {\bibinfo {author} {\bibfnamefont {F.}~\bibnamefont
  {Borgonovi}}\ and\ \bibinfo {author} {\bibfnamefont {D.~L.}\ \bibnamefont
  {Shepelyansky}},\ }\href {http://stacks.iop.org/0951-7715/8/i=5/a=013}
  {\bibfield  {journal} {\bibinfo  {journal} {Nonlinearity}\ }\textbf {\bibinfo
  {volume} {8}},\ \bibinfo {pages} {877} (\bibinfo {year} {1995})}\BibitemShut
  {NoStop}%
\bibitem [{\citenamefont {Wen-Lei}\ and\ \citenamefont
  {Quan-Lin}(2009)}]{WenLei09}%
  \BibitemOpen
  \bibfield  {author} {\bibinfo {author} {\bibfnamefont {Z.}~\bibnamefont
  {Wen-Lei}}\ and\ \bibinfo {author} {\bibfnamefont {J.}~\bibnamefont
  {Quan-Lin}},\ }\href {http://stacks.iop.org/0253-6102/51/i=3/a=17} {\bibfield
   {journal} {\bibinfo  {journal} {Comm. Theor. Phys.}\ }\textbf {\bibinfo
  {volume} {51}},\ \bibinfo {pages} {465} (\bibinfo {year} {2009})}\BibitemShut
  {NoStop}%
\bibitem [{\citenamefont {Wen-Lei}\ \emph {et~al.}(2010)\citenamefont
  {Wen-Lei}, \citenamefont {Quan-Lin},\ and\ \citenamefont {Bo}}]{WenLei10}%
  \BibitemOpen
  \bibfield  {author} {\bibinfo {author} {\bibfnamefont {Z.}~\bibnamefont
  {Wen-Lei}}, \bibinfo {author} {\bibfnamefont {J.}~\bibnamefont {Quan-Lin}}, \
  and\ \bibinfo {author} {\bibfnamefont {Z.}~\bibnamefont {Bo}},\ }\href
  {http://stacks.iop.org/0253-6102/54/i=2/a=09} {\bibfield  {journal} {\bibinfo
   {journal} {Comm. Theor. Phys.}\ }\textbf {\bibinfo {volume} {54}},\ \bibinfo
  {pages} {247} (\bibinfo {year} {2010})}\BibitemShut {NoStop}%
\bibitem [{\citenamefont {Gligori{\'{c}}}\ \emph {et~al.}(2011)\citenamefont
  {Gligori{\'{c}}}, \citenamefont {Bodyfelt},\ and\ \citenamefont
  {Flach}}]{GligoricBodyfeltFlach}%
  \BibitemOpen
  \bibfield  {author} {\bibinfo {author} {\bibfnamefont {G.}~\bibnamefont
  {Gligori{\'{c}}}}, \bibinfo {author} {\bibfnamefont {J.~D.}\ \bibnamefont
  {Bodyfelt}}, \ and\ \bibinfo {author} {\bibfnamefont {S.}~\bibnamefont
  {Flach}},\ }\href {\doibase 10.1209/0295-5075/96/30004} {\bibfield  {journal}
  {\bibinfo  {journal} {Europhys. Lett.}\ }\textbf {\bibinfo {volume} {96}},\
  \bibinfo {pages} {30004} (\bibinfo {year} {2011})}\BibitemShut {NoStop}%
\bibitem [{\citenamefont {Keser}\ \emph {et~al.}(2016)\citenamefont {Keser},
  \citenamefont {Ganeshan}, \citenamefont {Refael},\ and\ \citenamefont
  {Galitski}}]{KeserGaneshanRefaelGalitski}%
  \BibitemOpen
  \bibfield  {author} {\bibinfo {author} {\bibfnamefont {A.~C.}\ \bibnamefont
  {Keser}}, \bibinfo {author} {\bibfnamefont {S.}~\bibnamefont {Ganeshan}},
  \bibinfo {author} {\bibfnamefont {G.}~\bibnamefont {Refael}}, \ and\ \bibinfo
  {author} {\bibfnamefont {V.}~\bibnamefont {Galitski}},\ }\href {\doibase
  10.1103/PhysRevB.94.085120} {\bibfield  {journal} {\bibinfo  {journal} {Phys.
  Rev. B}\ }\textbf {\bibinfo {volume} {94}},\ \bibinfo {pages} {085120}
  (\bibinfo {year} {2016})}\BibitemShut {NoStop}%
\bibitem [{\citenamefont {Qin}\ \emph {et~al.}(2017)\citenamefont {Qin},
  \citenamefont {Andreanov}, \citenamefont {Park},\ and\ \citenamefont
  {Flach}}]{QinAndreanovParkFlach}%
  \BibitemOpen
  \bibfield  {author} {\bibinfo {author} {\bibfnamefont {P.}~\bibnamefont
  {Qin}}, \bibinfo {author} {\bibfnamefont {A.}~\bibnamefont {Andreanov}},
  \bibinfo {author} {\bibfnamefont {H.~C.}\ \bibnamefont {Park}}, \ and\
  \bibinfo {author} {\bibfnamefont {S.}~\bibnamefont {Flach}},\ }\href
  {https://doi.org/10.1038/srep41139} {\bibfield  {journal} {\bibinfo
  {journal} {Sci. Rep.}\ }\textbf {\bibinfo {volume} {7}},\ \bibinfo {pages}
  {41139} (\bibinfo {year} {2017})}\BibitemShut {NoStop}%
\bibitem [{\citenamefont {Toikka}\ and\ \citenamefont
  {Andreanov}(2019)}]{ToikkaAndreanov}%
  \BibitemOpen
  \bibfield  {author} {\bibinfo {author} {\bibfnamefont {L.~A.}\ \bibnamefont
  {Toikka}}\ and\ \bibinfo {author} {\bibfnamefont {A.}~\bibnamefont
  {Andreanov}},\ }\href {https://arxiv.org/abs/1901.09362} {\bibfield
  {journal} {\bibinfo  {journal} {arXiv:1901.09362}\ } (\bibinfo {year}
  {2019})}\BibitemShut {NoStop}%
\bibitem [{\citenamefont {Bloch}\ \emph {et~al.}(2008)\citenamefont {Bloch},
  \citenamefont {Dalibard},\ and\ \citenamefont
  {Zwerger}}]{BlochDalibardZwerger}%
  \BibitemOpen
  \bibfield  {author} {\bibinfo {author} {\bibfnamefont {I.}~\bibnamefont
  {Bloch}}, \bibinfo {author} {\bibfnamefont {J.}~\bibnamefont {Dalibard}}, \
  and\ \bibinfo {author} {\bibfnamefont {W.}~\bibnamefont {Zwerger}},\ }\href
  {\doibase 10.1103/RevModPhys.80.885} {\bibfield  {journal} {\bibinfo
  {journal} {Rev. Mod. Phys.}\ }\textbf {\bibinfo {volume} {80}},\ \bibinfo
  {pages} {885} (\bibinfo {year} {2008})}\BibitemShut {NoStop}%
\bibitem [{\citenamefont {Gadway}\ \emph {et~al.}(2013)\citenamefont {Gadway},
  \citenamefont {Reeves}, \citenamefont {Krinner},\ and\ \citenamefont
  {Schneble}}]{Gadway13}%
  \BibitemOpen
  \bibfield  {author} {\bibinfo {author} {\bibfnamefont {B.}~\bibnamefont
  {Gadway}}, \bibinfo {author} {\bibfnamefont {J.}~\bibnamefont {Reeves}},
  \bibinfo {author} {\bibfnamefont {L.}~\bibnamefont {Krinner}}, \ and\
  \bibinfo {author} {\bibfnamefont {D.}~\bibnamefont {Schneble}},\ }\href
  {\doibase 10.1103/PhysRevLett.110.190401} {\bibfield  {journal} {\bibinfo
  {journal} {Phys. Rev. Lett.}\ }\textbf {\bibinfo {volume} {110}},\ \bibinfo
  {pages} {190401} (\bibinfo {year} {2013})}\BibitemShut {NoStop}%
\bibitem [{\citenamefont {D’Alessio}\ and\ \citenamefont
  {Polkovnikov}(2013)}]{Dalessio}%
  \BibitemOpen
  \bibfield  {author} {\bibinfo {author} {\bibfnamefont {L.}~\bibnamefont
  {D’Alessio}}\ and\ \bibinfo {author} {\bibfnamefont {A.}~\bibnamefont
  {Polkovnikov}},\ }\href {\doibase 10.1016/j.aop.2013.02.011} {\bibfield
  {journal} {\bibinfo  {journal} {Annals of Physics}\ }\textbf {\bibinfo
  {volume} {333}},\ \bibinfo {pages} {19–33} (\bibinfo {year}
  {2013})}\BibitemShut {NoStop}%
\bibitem [{\citenamefont {Bukov}\ \emph {et~al.}(2015)\citenamefont {Bukov},
  \citenamefont {D’Alessio},\ and\ \citenamefont {Polkovnikov}}]{Bukov}%
  \BibitemOpen
  \bibfield  {author} {\bibinfo {author} {\bibfnamefont {M.}~\bibnamefont
  {Bukov}}, \bibinfo {author} {\bibfnamefont {L.}~\bibnamefont {D’Alessio}},
  \ and\ \bibinfo {author} {\bibfnamefont {A.}~\bibnamefont {Polkovnikov}},\
  }\href {\doibase 10.1080/00018732.2015.1055918} {\bibfield  {journal}
  {\bibinfo  {journal} {Advances in Physics}\ }\textbf {\bibinfo {volume}
  {64}},\ \bibinfo {pages} {139–226} (\bibinfo {year} {2015})}\BibitemShut
  {NoStop}%
\bibitem [{\citenamefont {Citro}\ \emph {et~al.}(2015)\citenamefont {Citro},
  \citenamefont {Dalla~Torre}, \citenamefont {D’Alessio}, \citenamefont
  {Polkovnikov}, \citenamefont {Babadi}, \citenamefont {Oka},\ and\
  \citenamefont {Demler}}]{Citro}%
  \BibitemOpen
  \bibfield  {author} {\bibinfo {author} {\bibfnamefont {R.}~\bibnamefont
  {Citro}}, \bibinfo {author} {\bibfnamefont {E.~G.}\ \bibnamefont
  {Dalla~Torre}}, \bibinfo {author} {\bibfnamefont {L.}~\bibnamefont
  {D’Alessio}}, \bibinfo {author} {\bibfnamefont {A.}~\bibnamefont
  {Polkovnikov}}, \bibinfo {author} {\bibfnamefont {M.}~\bibnamefont {Babadi}},
  \bibinfo {author} {\bibfnamefont {T.}~\bibnamefont {Oka}}, \ and\ \bibinfo
  {author} {\bibfnamefont {E.}~\bibnamefont {Demler}},\ }\href {\doibase
  10.1016/j.aop.2015.03.027} {\bibfield  {journal} {\bibinfo  {journal} {Annals
  of Physics}\ }\textbf {\bibinfo {volume} {360}},\ \bibinfo {pages}
  {694–710} (\bibinfo {year} {2015})}\BibitemShut {NoStop}%
\bibitem [{\citenamefont {Rajak}\ \emph {et~al.}(2019)\citenamefont {Rajak},
  \citenamefont {Dana},\ and\ \citenamefont {Dalla~Torre}}]{Rajak}%
  \BibitemOpen
  \bibfield  {author} {\bibinfo {author} {\bibfnamefont {A.}~\bibnamefont
  {Rajak}}, \bibinfo {author} {\bibfnamefont {I.}~\bibnamefont {Dana}}, \ and\
  \bibinfo {author} {\bibfnamefont {E.~G.}\ \bibnamefont {Dalla~Torre}},\
  }\href {\doibase 10.1103/PhysRevB.100.100302} {\bibfield  {journal} {\bibinfo
   {journal} {Phys. Rev. B}\ }\textbf {\bibinfo {volume} {100}},\ \bibinfo
  {pages} {100302} (\bibinfo {year} {2019})}\BibitemShut {NoStop}%
\bibitem [{\citenamefont {Giamarchi}(2003)}]{TG}%
  \BibitemOpen
  \bibfield  {author} {\bibinfo {author} {\bibfnamefont {T.}~\bibnamefont
  {Giamarchi}},\ }\href@noop {} {\emph {\bibinfo {title} {Quantum Physics in
  One Dimension}}},\ International Series of Monographs on Physics\ (\bibinfo
  {publisher} {Clarendon Press},\ \bibinfo {year} {2003})\BibitemShut {NoStop}%
\bibitem [{\citenamefont {Gogolin}\ \emph {et~al.}(2004)\citenamefont
  {Gogolin}, \citenamefont {Nersesyan},\ and\ \citenamefont
  {Tsvelik}}]{GogolinNerseyanTsvelik}%
  \BibitemOpen
  \bibfield  {author} {\bibinfo {author} {\bibfnamefont {A.}~\bibnamefont
  {Gogolin}}, \bibinfo {author} {\bibfnamefont {A.}~\bibnamefont {Nersesyan}},
  \ and\ \bibinfo {author} {\bibfnamefont {A.}~\bibnamefont {Tsvelik}},\ }\href
  {https://books.google.com/books?id=BZDfFIpCoaAC} {\emph {\bibinfo {title}
  {Bosonization and Strongly Correlated Systems}}}\ (\bibinfo  {publisher}
  {Cambridge University Press},\ \bibinfo {year} {2004})\BibitemShut {NoStop}%
\bibitem [{\citenamefont {Bertini}\ \emph {et~al.}(2016)\citenamefont
  {Bertini}, \citenamefont {Collura}, \citenamefont {De~Nardis},\ and\
  \citenamefont {Fagotti}}]{BertiniColluraDenardisFagotti}%
  \BibitemOpen
  \bibfield  {author} {\bibinfo {author} {\bibfnamefont {B.}~\bibnamefont
  {Bertini}}, \bibinfo {author} {\bibfnamefont {M.}~\bibnamefont {Collura}},
  \bibinfo {author} {\bibfnamefont {J.}~\bibnamefont {De~Nardis}}, \ and\
  \bibinfo {author} {\bibfnamefont {M.}~\bibnamefont {Fagotti}},\ }\href
  {\doibase 10.1103/PhysRevLett.117.207201} {\bibfield  {journal} {\bibinfo
  {journal} {Phys. Rev. Lett.}\ }\textbf {\bibinfo {volume} {117}},\ \bibinfo
  {pages} {207201} (\bibinfo {year} {2016})}\BibitemShut {NoStop}%
\bibitem [{\citenamefont {Castro-Alvaredo}\ \emph {et~al.}(2016)\citenamefont
  {Castro-Alvaredo}, \citenamefont {Doyon},\ and\ \citenamefont
  {Yoshimura}}]{CastroDoyon}%
  \BibitemOpen
  \bibfield  {author} {\bibinfo {author} {\bibfnamefont {O.~A.}\ \bibnamefont
  {Castro-Alvaredo}}, \bibinfo {author} {\bibfnamefont {B.}~\bibnamefont
  {Doyon}}, \ and\ \bibinfo {author} {\bibfnamefont {T.}~\bibnamefont
  {Yoshimura}},\ }\href {\doibase 10.1103/PhysRevX.6.041065} {\bibfield
  {journal} {\bibinfo  {journal} {Phys. Rev. X}\ }\textbf {\bibinfo {volume}
  {6}},\ \bibinfo {pages} {041065} (\bibinfo {year} {2016})}\BibitemShut
  {NoStop}%
\bibitem [{\citenamefont {Lieb}\ and\ \citenamefont
  {Liniger}(1963)}]{LiebLiniger1}%
  \BibitemOpen
  \bibfield  {author} {\bibinfo {author} {\bibfnamefont {E.~H.}\ \bibnamefont
  {Lieb}}\ and\ \bibinfo {author} {\bibfnamefont {W.}~\bibnamefont {Liniger}},\
  }\href {\doibase 10.1103/PhysRev.130.1605} {\bibfield  {journal} {\bibinfo
  {journal} {Phys. Rev.}\ }\textbf {\bibinfo {volume} {130}},\ \bibinfo {pages}
  {1605} (\bibinfo {year} {1963})}\BibitemShut {NoStop}%
\bibitem [{\citenamefont {Lieb}(1963)}]{LiebLiniger2}%
  \BibitemOpen
  \bibfield  {author} {\bibinfo {author} {\bibfnamefont {E.~H.}\ \bibnamefont
  {Lieb}},\ }\href {\doibase 10.1103/PhysRev.130.1616} {\bibfield  {journal}
  {\bibinfo  {journal} {Phys. Rev.}\ }\textbf {\bibinfo {volume} {130}},\
  \bibinfo {pages} {1616} (\bibinfo {year} {1963})}\BibitemShut {NoStop}%
\bibitem [{\citenamefont {Olshanii}(1998)}]{Olshanii}%
  \BibitemOpen
  \bibfield  {author} {\bibinfo {author} {\bibfnamefont {M.}~\bibnamefont
  {Olshanii}},\ }\href {\doibase 10.1103/PhysRevLett.81.938} {\bibfield
  {journal} {\bibinfo  {journal} {Phys. Rev. Lett.}\ }\textbf {\bibinfo
  {volume} {81}},\ \bibinfo {pages} {938} (\bibinfo {year} {1998})}\BibitemShut
  {NoStop}%
\bibitem [{\citenamefont {Dunjko}\ \emph {et~al.}(2001)\citenamefont {Dunjko},
  \citenamefont {Lorent},\ and\ \citenamefont
  {Olshanii}}]{DunjkoLorentOlshanii}%
  \BibitemOpen
  \bibfield  {author} {\bibinfo {author} {\bibfnamefont {V.}~\bibnamefont
  {Dunjko}}, \bibinfo {author} {\bibfnamefont {V.}~\bibnamefont {Lorent}}, \
  and\ \bibinfo {author} {\bibfnamefont {M.}~\bibnamefont {Olshanii}},\ }\href
  {\doibase 10.1103/PhysRevLett.86.5413} {\bibfield  {journal} {\bibinfo
  {journal} {Phys. Rev. Lett.}\ }\textbf {\bibinfo {volume} {86}},\ \bibinfo
  {pages} {5413} (\bibinfo {year} {2001})}\BibitemShut {NoStop}%
\bibitem [{\citenamefont {{Gaudin}}\ and\ \citenamefont
  {{Caux}}(2014)}]{Gaudin}%
  \BibitemOpen
  \bibfield  {author} {\bibinfo {author} {\bibfnamefont {M.}~\bibnamefont
  {{Gaudin}}}\ and\ \bibinfo {author} {\bibfnamefont {J.-S.}\ \bibnamefont
  {{Caux}}},\ }\href {http://adsabs.harvard.edu/abs/2014bewa.book.....G} {\emph
  {\bibinfo {title} {The Bethe Wavefunction}}}\ (\bibinfo  {publisher}
  {Cambridge University Press},\ \bibinfo {year} {2014})\BibitemShut {NoStop}%
\bibitem [{\citenamefont {{Korepin}}\ \emph {et~al.}(1993)\citenamefont
  {{Korepin}}, \citenamefont {{Bogoliubov}},\ and\ \citenamefont
  {{Izergin}}}]{KorepinBook}%
  \BibitemOpen
  \bibfield  {author} {\bibinfo {author} {\bibfnamefont {V.~E.}\ \bibnamefont
  {{Korepin}}}, \bibinfo {author} {\bibfnamefont {N.~M.}\ \bibnamefont
  {{Bogoliubov}}}, \ and\ \bibinfo {author} {\bibfnamefont {A.~G.}\
  \bibnamefont {{Izergin}}},\ }\href@noop {} {\emph {\bibinfo {title} {Quantum
  Inverse Scattering Method and Correlation Functions}}}\ (\bibinfo
  {publisher} {Cambridge University Press},\ \bibinfo {year} {1993})\ p.\
  \bibinfo {pages} {575}\BibitemShut {NoStop}%
\bibitem [{\citenamefont {Kormos}\ \emph {et~al.}(2013)\citenamefont {Kormos},
  \citenamefont {Shashi}, \citenamefont {Chou}, \citenamefont {Caux},\ and\
  \citenamefont {Imambekov}}]{KormosShashiChouCauxImambekov}%
  \BibitemOpen
  \bibfield  {author} {\bibinfo {author} {\bibfnamefont {M.}~\bibnamefont
  {Kormos}}, \bibinfo {author} {\bibfnamefont {A.}~\bibnamefont {Shashi}},
  \bibinfo {author} {\bibfnamefont {Y.-Z.}\ \bibnamefont {Chou}}, \bibinfo
  {author} {\bibfnamefont {J.-S.}\ \bibnamefont {Caux}}, \ and\ \bibinfo
  {author} {\bibfnamefont {A.}~\bibnamefont {Imambekov}},\ }\href {\doibase
  10.1103/PhysRevB.88.205131} {\bibfield  {journal} {\bibinfo  {journal} {Phys.
  Rev. B}\ }\textbf {\bibinfo {volume} {88}},\ \bibinfo {pages} {205131}
  (\bibinfo {year} {2013})}\BibitemShut {NoStop}%
\bibitem [{\citenamefont {De~Nardis}\ \emph {et~al.}(2014)\citenamefont
  {De~Nardis}, \citenamefont {Wouters}, \citenamefont {Brockmann},\ and\
  \citenamefont {Caux}}]{DeNardiWoutersBrockmanCaux}%
  \BibitemOpen
  \bibfield  {author} {\bibinfo {author} {\bibfnamefont {J.}~\bibnamefont
  {De~Nardis}}, \bibinfo {author} {\bibfnamefont {B.}~\bibnamefont {Wouters}},
  \bibinfo {author} {\bibfnamefont {M.}~\bibnamefont {Brockmann}}, \ and\
  \bibinfo {author} {\bibfnamefont {J.-S.}\ \bibnamefont {Caux}},\ }\href
  {\doibase 10.1103/PhysRevA.89.033601} {\bibfield  {journal} {\bibinfo
  {journal} {Phys. Rev. A}\ }\textbf {\bibinfo {volume} {89}},\ \bibinfo
  {pages} {033601} (\bibinfo {year} {2014})}\BibitemShut {NoStop}%
\bibitem [{\citenamefont {Caux}\ and\ \citenamefont {Konik}(2012)}]{CauxKonik}%
  \BibitemOpen
  \bibfield  {author} {\bibinfo {author} {\bibfnamefont {J.-S.}\ \bibnamefont
  {Caux}}\ and\ \bibinfo {author} {\bibfnamefont {R.~M.}\ \bibnamefont
  {Konik}},\ }\href {\doibase 10.1103/PhysRevLett.109.175301} {\bibfield
  {journal} {\bibinfo  {journal} {Phys. Rev. Lett.}\ }\textbf {\bibinfo
  {volume} {109}},\ \bibinfo {pages} {175301} (\bibinfo {year}
  {2012})}\BibitemShut {NoStop}%
\bibitem [{\citenamefont {Tonks}(1936)}]{Tonks}%
  \BibitemOpen
  \bibfield  {author} {\bibinfo {author} {\bibfnamefont {L.}~\bibnamefont
  {Tonks}},\ }\href {\doibase 10.1103/PhysRev.50.955} {\bibfield  {journal}
  {\bibinfo  {journal} {Phys. Rev.}\ }\textbf {\bibinfo {volume} {50}},\
  \bibinfo {pages} {955} (\bibinfo {year} {1936})}\BibitemShut {NoStop}%
\bibitem [{\citenamefont {{Girardeau}}(1960)}]{Girardeau}%
  \BibitemOpen
  \bibfield  {author} {\bibinfo {author} {\bibfnamefont {M.}~\bibnamefont
  {{Girardeau}}},\ }\href {\doibase 10.1063/1.1703687} {\bibfield  {journal}
  {\bibinfo  {journal} {J. Math. Phys.}\ }\textbf {\bibinfo {volume} {1}},\
  \bibinfo {pages} {516} (\bibinfo {year} {1960})}\BibitemShut {NoStop}%
\bibitem [{\citenamefont {{Yukalov}}\ and\ \citenamefont
  {{Girardeau}}(2005)}]{YukalovGurardeau}%
  \BibitemOpen
  \bibfield  {author} {\bibinfo {author} {\bibfnamefont {V.~I.}\ \bibnamefont
  {{Yukalov}}}\ and\ \bibinfo {author} {\bibfnamefont {M.~D.}\ \bibnamefont
  {{Girardeau}}},\ }\href {\doibase 10.1002/lapl.200510011} {\bibfield
  {journal} {\bibinfo  {journal} {Laser Phys. Lett.}\ }\textbf {\bibinfo
  {volume} {2}},\ \bibinfo {pages} {375} (\bibinfo {year} {2005})}\BibitemShut
  {NoStop}%
\bibitem [{\citenamefont {{Pezer}}\ and\ \citenamefont
  {{Buljan}}(2007)}]{Pezer}%
  \BibitemOpen
  \bibfield  {author} {\bibinfo {author} {\bibfnamefont {R.}~\bibnamefont
  {{Pezer}}}\ and\ \bibinfo {author} {\bibfnamefont {H.}~\bibnamefont
  {{Buljan}}},\ }\href {\doibase 10.1103/PhysRevLett.98.240403} {\bibfield
  {journal} {\bibinfo  {journal} {Phys. Rev. Lett.}\ }\textbf {\bibinfo
  {volume} {98}},\ \bibinfo {eid} {240403} (\bibinfo {year}
  {2007})}\BibitemShut {NoStop}%
\bibitem [{\citenamefont {Izergin}\ and\ \citenamefont
  {Korepin}(1986)}]{IzereginKorepin}%
  \BibitemOpen
  \bibfield  {author} {\bibinfo {author} {\bibfnamefont {A.}~\bibnamefont
  {Izergin}}\ and\ \bibinfo {author} {\bibfnamefont {V.}~\bibnamefont
  {Korepin}},\ }\href {\doibase https://doi.org/10.1007/BF01095102} {\bibfield
  {journal} {\bibinfo  {journal} {J. Math. Sci.}\ }\textbf {\bibinfo {volume}
  {34}},\ \bibinfo {pages} {1933} (\bibinfo {year} {1986})}\BibitemShut
  {NoStop}%
\bibitem [{\citenamefont {Haldane}(1981)}]{haldane}%
  \BibitemOpen
  \bibfield  {author} {\bibinfo {author} {\bibfnamefont {F.~D.~M.}\
  \bibnamefont {Haldane}},\ }\href {\doibase 10.1103/PhysRevLett.47.1840}
  {\bibfield  {journal} {\bibinfo  {journal} {Phys. Rev. Lett.}\ }\textbf
  {\bibinfo {volume} {47}},\ \bibinfo {pages} {1840} (\bibinfo {year}
  {1981})}\BibitemShut {NoStop}%
\bibitem [{\citenamefont {{Mattis}}\ and\ \citenamefont
  {{Lieb}}(1965)}]{MattisLieb}%
  \BibitemOpen
  \bibfield  {author} {\bibinfo {author} {\bibfnamefont {D.~C.}\ \bibnamefont
  {{Mattis}}}\ and\ \bibinfo {author} {\bibfnamefont {E.~H.}\ \bibnamefont
  {{Lieb}}},\ }\href {\doibase 10.1063/1.1704281} {\bibfield  {journal}
  {\bibinfo  {journal} {J. Math. Phys.}\ }\textbf {\bibinfo {volume} {6}},\
  \bibinfo {pages} {304} (\bibinfo {year} {1965})}\BibitemShut {NoStop}%
\bibitem [{\citenamefont {{Rozhkov}}(2005)}]{Rozhkov}%
  \BibitemOpen
  \bibfield  {author} {\bibinfo {author} {\bibfnamefont {A.~V.}\ \bibnamefont
  {{Rozhkov}}},\ }\href {\doibase 10.1140/epjb/e2005-00312-3} {\bibfield
  {journal} {\bibinfo  {journal} {Euro. Phys. J. B}\ }\textbf {\bibinfo
  {volume} {47}},\ \bibinfo {pages} {193} (\bibinfo {year} {2005})}\BibitemShut
  {NoStop}%
\bibitem [{\citenamefont {{Cazalilla}}(2004)}]{Cazalilla}%
  \BibitemOpen
  \bibfield  {author} {\bibinfo {author} {\bibfnamefont {M.~A.}\ \bibnamefont
  {{Cazalilla}}},\ }\href {\doibase 10.1088/0953-4075/37/7/051} {\bibfield
  {journal} {\bibinfo  {journal} {J. Phys. B}\ }\textbf {\bibinfo {volume}
  {37}},\ \bibinfo {pages} {S1} (\bibinfo {year} {2004})}\BibitemShut {NoStop}%
\bibitem [{Sup({\natexlab{a}})}]{SupplementS2}%
  \BibitemOpen
  \bibinfo {note} {See the supplemental material, section 2 for
  details.}\BibitemShut {Stop}%
\bibitem [{\citenamefont {Grempel}\ \emph {et~al.}(1982)\citenamefont
  {Grempel}, \citenamefont {Fishman},\ and\ \citenamefont
  {Prange}}]{GrempelFishmanPrangeLinear}%
  \BibitemOpen
  \bibfield  {author} {\bibinfo {author} {\bibfnamefont {D.~R.}\ \bibnamefont
  {Grempel}}, \bibinfo {author} {\bibfnamefont {S.}~\bibnamefont {Fishman}}, \
  and\ \bibinfo {author} {\bibfnamefont {R.~E.}\ \bibnamefont {Prange}},\
  }\href {\doibase 10.1103/PhysRevLett.49.833} {\bibfield  {journal} {\bibinfo
  {journal} {Phys. Rev. Lett.}\ }\textbf {\bibinfo {volume} {49}},\ \bibinfo
  {pages} {833} (\bibinfo {year} {1982})}\BibitemShut {NoStop}%
\bibitem [{\citenamefont {Berry}(1984)}]{Berry}%
  \BibitemOpen
  \bibfield  {author} {\bibinfo {author} {\bibfnamefont {M.}~\bibnamefont
  {Berry}},\ }\href {\doibase doi.org/10.1016/0167-2789(84)90185-4} {\bibfield
  {journal} {\bibinfo  {journal} {Physica D}\ }\textbf {\bibinfo {volume}
  {10}},\ \bibinfo {pages} {369 } (\bibinfo {year} {1984})}\BibitemShut
  {NoStop}%
\bibitem [{\citenamefont {Imambekov}\ \emph {et~al.}(2012)\citenamefont
  {Imambekov}, \citenamefont {Schmidt},\ and\ \citenamefont
  {Glazman}}]{ImamGlazRMP}%
  \BibitemOpen
  \bibfield  {author} {\bibinfo {author} {\bibfnamefont {A.}~\bibnamefont
  {Imambekov}}, \bibinfo {author} {\bibfnamefont {T.~L.}\ \bibnamefont
  {Schmidt}}, \ and\ \bibinfo {author} {\bibfnamefont {L.~I.}\ \bibnamefont
  {Glazman}},\ }\href {\doibase 10.1103/RevModPhys.84.1253} {\bibfield
  {journal} {\bibinfo  {journal} {Rev. Mod. Phys.}\ }\textbf {\bibinfo {volume}
  {84}},\ \bibinfo {pages} {1253} (\bibinfo {year} {2012})}\BibitemShut
  {NoStop}%
\bibitem [{\citenamefont {Pereira}\ \emph {et~al.}(2008)\citenamefont
  {Pereira}, \citenamefont {White},\ and\ \citenamefont {Affleck}}]{Pereira}%
  \BibitemOpen
  \bibfield  {author} {\bibinfo {author} {\bibfnamefont {R.~G.}\ \bibnamefont
  {Pereira}}, \bibinfo {author} {\bibfnamefont {S.~R.}\ \bibnamefont {White}},
  \ and\ \bibinfo {author} {\bibfnamefont {I.}~\bibnamefont {Affleck}},\ }\href
  {\doibase 10.1103/PhysRevLett.100.027206} {\bibfield  {journal} {\bibinfo
  {journal} {Phys. Rev. Lett.}\ }\textbf {\bibinfo {volume} {100}},\ \bibinfo
  {pages} {027206} (\bibinfo {year} {2008})}\BibitemShut {NoStop}%
\bibitem [{Sup({\natexlab{b}})}]{SupplementS4}%
  \BibitemOpen
  \bibinfo {note} {See the supplemental material, section 4 for
  details.}\BibitemShut {Stop}%
\bibitem [{\citenamefont {Doyon}\ \emph {et~al.}(2017)\citenamefont {Doyon},
  \citenamefont {Dubail}, \citenamefont {Konik},\ and\ \citenamefont
  {Yoshimura}}]{DoyonDubailKonikYoshimura}%
  \BibitemOpen
  \bibfield  {author} {\bibinfo {author} {\bibfnamefont {B.}~\bibnamefont
  {Doyon}}, \bibinfo {author} {\bibfnamefont {J.}~\bibnamefont {Dubail}},
  \bibinfo {author} {\bibfnamefont {R.}~\bibnamefont {Konik}}, \ and\ \bibinfo
  {author} {\bibfnamefont {T.}~\bibnamefont {Yoshimura}},\ }\href {\doibase
  10.1103/PhysRevLett.119.195301} {\bibfield  {journal} {\bibinfo  {journal}
  {Phys. Rev. Lett.}\ }\textbf {\bibinfo {volume} {119}},\ \bibinfo {pages}
  {195301} (\bibinfo {year} {2017})}\BibitemShut {NoStop}%
\bibitem [{\citenamefont {Wigner}(1932)}]{Wigner}%
  \BibitemOpen
  \bibfield  {author} {\bibinfo {author} {\bibfnamefont {E.}~\bibnamefont
  {Wigner}},\ }\href {\doibase 10.1103/PhysRev.40.749} {\bibfield  {journal}
  {\bibinfo  {journal} {Phys. Rev.}\ }\textbf {\bibinfo {volume} {40}},\
  \bibinfo {pages} {749} (\bibinfo {year} {1932})}\BibitemShut {NoStop}%
\bibitem [{\citenamefont {Doyon}\ and\ \citenamefont
  {Yoshimura}(2017)}]{DoyonYoshimura}%
  \BibitemOpen
  \bibfield  {author} {\bibinfo {author} {\bibfnamefont {B.}~\bibnamefont
  {Doyon}}\ and\ \bibinfo {author} {\bibfnamefont {T.}~\bibnamefont
  {Yoshimura}},\ }\href {\doibase 10.21468/SciPostPhys.2.2.014} {\bibfield
  {journal} {\bibinfo  {journal} {SciPost Phys.}\ }\textbf {\bibinfo {volume}
  {2}},\ \bibinfo {pages} {014} (\bibinfo {year} {2017})}\BibitemShut {NoStop}%
\bibitem [{Sup({\natexlab{c}})}]{SupplementS1}%
  \BibitemOpen
  \bibinfo {note} {See the supplemental material, section 1 for
  details.}\BibitemShut {Stop}%
\bibitem [{\citenamefont {Fishman}\ \emph {et~al.}(1984)\citenamefont
  {Fishman}, \citenamefont {Grempel},\ and\ \citenamefont
  {Prange}}]{Fishman84}%
  \BibitemOpen
  \bibfield  {author} {\bibinfo {author} {\bibfnamefont {S.}~\bibnamefont
  {Fishman}}, \bibinfo {author} {\bibfnamefont {D.~R.}\ \bibnamefont
  {Grempel}}, \ and\ \bibinfo {author} {\bibfnamefont {R.~E.}\ \bibnamefont
  {Prange}},\ }\href {\doibase 10.1103/PhysRevB.29.4272} {\bibfield  {journal}
  {\bibinfo  {journal} {Phys. Rev. B}\ }\textbf {\bibinfo {volume} {29}},\
  \bibinfo {pages} {4272} (\bibinfo {year} {1984})}\BibitemShut {NoStop}%
\bibitem [{\citenamefont {Prange}\ \emph {et~al.}(1984)\citenamefont {Prange},
  \citenamefont {Grempel},\ and\ \citenamefont {Fishman}}]{Prange84}%
  \BibitemOpen
  \bibfield  {author} {\bibinfo {author} {\bibfnamefont {R.~E.}\ \bibnamefont
  {Prange}}, \bibinfo {author} {\bibfnamefont {D.~R.}\ \bibnamefont {Grempel}},
  \ and\ \bibinfo {author} {\bibfnamefont {S.}~\bibnamefont {Fishman}},\ }\href
  {\doibase 10.1103/PhysRevB.29.6500} {\bibfield  {journal} {\bibinfo
  {journal} {Phys. Rev. B}\ }\textbf {\bibinfo {volume} {29}},\ \bibinfo
  {pages} {6500} (\bibinfo {year} {1984})}\BibitemShut {NoStop}%
\bibitem [{\citenamefont {Lima}\ and\ \citenamefont
  {Shepelyansky}(1991)}]{Lima91}%
  \BibitemOpen
  \bibfield  {author} {\bibinfo {author} {\bibfnamefont {R.}~\bibnamefont
  {Lima}}\ and\ \bibinfo {author} {\bibfnamefont {D.}~\bibnamefont
  {Shepelyansky}},\ }\href {\doibase 10.1103/PhysRevLett.67.1377} {\bibfield
  {journal} {\bibinfo  {journal} {Phys. Rev. Lett.}\ }\textbf {\bibinfo
  {volume} {67}},\ \bibinfo {pages} {1377} (\bibinfo {year}
  {1991})}\BibitemShut {NoStop}%
\bibitem [{\citenamefont {Borgonovi}\ and\ \citenamefont
  {Shepelyansky}(1997)}]{Borgonovi97}%
  \BibitemOpen
  \bibfield  {author} {\bibinfo {author} {\bibfnamefont {F.}~\bibnamefont
  {Borgonovi}}\ and\ \bibinfo {author} {\bibfnamefont {D.}~\bibnamefont
  {Shepelyansky}},\ }\href {\doibase doi.org/10.1016/S0167-2789(97)00155-3}
  {\bibfield  {journal} {\bibinfo  {journal} {Physica D}\ }\textbf {\bibinfo
  {volume} {109}},\ \bibinfo {pages} {24 } (\bibinfo {year}
  {1997})}\BibitemShut {NoStop}%
\bibitem [{\citenamefont {Bulchandani}\ \emph {et~al.}(2017)\citenamefont
  {Bulchandani}, \citenamefont {Vasseur}, \citenamefont {Karrasch},\ and\
  \citenamefont {Moore}}]{BulchandaniVasseur}%
  \BibitemOpen
  \bibfield  {author} {\bibinfo {author} {\bibfnamefont {V.~B.}\ \bibnamefont
  {Bulchandani}}, \bibinfo {author} {\bibfnamefont {R.}~\bibnamefont
  {Vasseur}}, \bibinfo {author} {\bibfnamefont {C.}~\bibnamefont {Karrasch}}, \
  and\ \bibinfo {author} {\bibfnamefont {J.~E.}\ \bibnamefont {Moore}},\ }\href
  {\doibase 10.1103/PhysRevLett.119.220604} {\bibfield  {journal} {\bibinfo
  {journal} {Phys. Rev. Lett.}\ }\textbf {\bibinfo {volume} {119}},\ \bibinfo
  {pages} {220604} (\bibinfo {year} {2017})}\BibitemShut {NoStop}%
\bibitem [{\citenamefont {{Panfil}}\ and\ \citenamefont
  {{Pawe{\l}czyk}}(2019)}]{PanfilPawelczyk}%
  \BibitemOpen
  \bibfield  {author} {\bibinfo {author} {\bibfnamefont {M.}~\bibnamefont
  {{Panfil}}}\ and\ \bibinfo {author} {\bibfnamefont {J.}~\bibnamefont
  {{Pawe{\l}czyk}}},\ }\href {https://arxiv.org/abs/1905.06257} {\bibfield
  {journal} {\bibinfo  {journal} {arXiv:1905.06257}\ } (\bibinfo {year}
  {2019})}\BibitemShut {NoStop}%
\bibitem [{\citenamefont {Chirikov}(1979)}]{Chirikov79}%
  \BibitemOpen
  \bibfield  {author} {\bibinfo {author} {\bibfnamefont {B.~V.}\ \bibnamefont
  {Chirikov}},\ }\href {\doibase
  http://dx.doi.org/10.1016/0370-1573(79)90023-1} {\bibfield  {journal}
  {\bibinfo  {journal} {Phys. Rep.}\ }\textbf {\bibinfo {volume} {52}},\
  \bibinfo {pages} {263 } (\bibinfo {year} {1979})}\BibitemShut {NoStop}%
\bibitem [{Sup({\natexlab{d}})}]{SupplementS3}%
  \BibitemOpen
  \bibinfo {note} {See the supplemental material, section 3 for
  details.}\BibitemShut {Stop}%
\end{thebibliography}%

\beginsupplement

	\section{1. TG  Gas \& Wigner Function}
	In the main text we provided a simple argument for the dynamical localization of the TG gas based on the exact wavefunction of the model. One can also examine it from a different perspective which can be more readily generalised to other cases. Through Fermi-Bose mapping we can write the the TG Hamiltonian and kick as\cite{Tonks, Girardeau}
	\begin{eqnarray}
	H_\text{TG}=\int\mathrm{d}x\Psi^\dag(x)\left[-\frac{\partial_x^2}{2m}\right]\Psi(x), ~~H_\text{K}=\int\mathrm{d}x\,V\cos{(qx)}\Psi^\dag(x)\Psi(x)
	\end{eqnarray}
	We then study the time evolution of the Wigner function \cite{Wigner}
	\begin{eqnarray}
	n(x,\lambda)=\int\mathrm{d}y\,e^{i\lambda y}\left<\Psi^\dag(x+y/2)\Psi(x-y/2)\right>
	\end{eqnarray}
	Which follows a two step pattern.  Between the kicks, using the Heisenberg equations of motion this evolves according to 
	\begin{eqnarray}
	\left[\partial_t+v_\text{eff}(\lambda)\partial_x\right]n(x,\lambda,t)=0
	\end{eqnarray}
	where $v_\text{eff}(\lambda)=\lambda/m$. The solution of which is simply
	\begin{eqnarray}\label{freeEvo}
	n(x,\lambda,t+T^-)=n(x-\lambda T/m,\lambda, t).
	\end{eqnarray}
	Meanwhile, through the kicks we can use the fact that
	\begin{eqnarray}
	e^{iH_\text{K}}\Psi^\dag(x)e^{-iH_\text{K}}=e^{-iV\cos{qx}}\Psi^\dag(x)
	\end{eqnarray}
	from which we get that 
	\begin{eqnarray}\label{nKick}
	\tilde{n}(x,z,t+T^+)=e^{2iV\sin{(qz/2)}\sin{(qx)}}\tilde{n}(x,z,t+T^-)
	\end{eqnarray}
	where $\tilde{n}(x,z,t)$ is the Fourier transform of $n$ with respect to $\lambda$.  Repeated application of this two step evolution provides the full evolution of the TG Wigner function. This can be better achieved numerically by working in the Fourier space of $x$, denoting the Fourier transform with respect to  $x $ by $\bar{n}(p,\lambda,t)$ the free evolution is given by 
	\begin{eqnarray}
	\bar{n}(p,\lambda,t+T^-)=e^{-ip v_\text{eff}T}\bar{n}(p,\lambda,t).
	\end{eqnarray} 
	
	Having determined $n(x,\lambda,t)$ at any time, it can then be used to find  the density and momentum distribution function via integration over $x$ or $\lambda$ respectively
	\begin{eqnarray}
	n(\lambda,t)=\int\mathrm{d}x\, n(x,\lambda,t)\\
	\rho(x,t)=\int\frac{\mathrm{d}\lambda}{2\pi}\, n(x,\lambda,t)
	\end{eqnarray}
	The energy of the system is therefore given by 
	\begin{eqnarray}
	E(t)=\int\frac{\mathrm{d}\lambda}{2\pi}\,\left[\frac{\lambda^2}{2m}\right] n(\lambda,t)
	\end{eqnarray}
	This method allows one easily investigate the effects of different initial conditions. An initial trapping potential or finite temperature state could also be considered and moreover it can  be generalised away from the TG limit to give the GHD approach presented in the text \cite{DoyonDubailKonikYoshimura}. 
	
	The same analysis can be carried out in the free boson case with the exact same evolution. The difference between the cases only arising in the choice of initial condition, for the TG gas a natural choice is the Fermi-Dira distribution $n(x,\lambda,0)=\Theta(\lambda_F-\lambda)-\Theta(-\lambda_F-\lambda)$ where $\lambda_F$ is the Fermi momentum and $\Theta(x)$ a Heaviside function. 
	
	We conclude this section by noting an interesting relation to the classical kicked rotor system. Using the Heisenberg equations of motion for $n(x,\lambda,t)$, the effect of the kick can be determined via the solution of 
	\begin{eqnarray}
	\partial_tn(x,\lambda,t)=\sum_{j=0}^\infty \delta(t-jT)V\sin{(qx)}\left[n(x,\lambda+q/2,t)-n(x,\lambda-q/2,t)\right].
	\end{eqnarray}
	For a sufficiently smooth $n(x,\lambda,t)$ and small $q$ the right hand side can be expanded in a Taylor series. Retaining only the leading term in this sequence we have that the effect of the kick becomes 
	\begin{eqnarray}
	n(x,\lambda,t+T^+)=n(x,\lambda+Vq\sin(qx),t+T^-).
	\end{eqnarray}
	Combined with \eqref{freeEvo} we recover exactly the Chirikov standard map\cite{Chirikov79}. Such an approximation breaks down at zero temperature when the initial state is a Fermi function but may be appropriate at higher temperature.

	%%%%%%%%%%%%%%%%

	%%%%%%%%%%%%
	
	\section{2. Kicked Luttinger Liquid}
	Here we determine the Floquet Hamiltonian for the Kicked Luttinger Liquid. We employ the  bosonic form the Hamiltonian, which is
	\begin{eqnarray}\label{HLuttS}
	H_\text{Lutt}=\frac{v_s}{2\pi}\int\mathrm{d}x\, \frac{1}{K}\left[\nabla \phi(x)\right]^2+K\left[\nabla \theta(x)\right]^2
	\end{eqnarray}
	where $\phi(x)$ and $\theta(x)$ are bosonic fields related to the density and current of the system and satisfy 
	\begin{eqnarray}
	\left[\phi(x), \theta(y)\right]=i\pi\,\text{sgn}(y-x)~~
	\left[\phi(x), \nabla\theta(y)\right]=i\pi\delta(y-x)\\
	\left[\nabla\phi(x), \theta(y)\right]=-i\pi\delta(y-x)~~
	\left[\nabla\phi(x), \nabla\theta(y)\right]=i\pi\partial_x[\delta(y-x)].
	\end{eqnarray}
	The parameters, $v_s$ and $K$ incorporate the parameters of the Lieb Linger. For $K=1$, $v_s=v_F$ the Hamiltonian corresponds to the low energy of the TG gas. 
	The relation to the fermioninc description in the main text is given by \cite{TG}
	\begin{eqnarray}
	\nabla \phi(x)=-\pi\left[\rho_+(x)+\rho_-(x)\right],~\nabla\theta(x)=\pi \left[\rho_+(x)-\rho_-(x)\right]
	\end{eqnarray}
	In the bosonic language the density operator is just $-\frac{1}{\pi}\nabla \phi(x)$ so the kicking term is 
	\begin{eqnarray}\label{Hkick}
	H_\text{K}=\sum_{j=-\infty}^\infty\delta(t-jT)\int \mathrm{d}x\,V\cos{(qx)}\left[-\frac{1}{\pi}\nabla\phi(x)\right].
	\end{eqnarray}
	The Floquet Hamiltonian is defined by 
	\begin{eqnarray}
	e^{-iH_\text{F}}=e^{-i\frac{v_sT}{2\pi}\int\mathrm{d}x\, \frac{1}{K}\left[\nabla \phi(x)\right]^2+K\left[\nabla \theta(x)\right]^2}e^{i\frac{V}{\pi}\int \mathrm{d}x\,\cos{(qx)}\nabla\phi(x)}
	\end{eqnarray}
	We can combine these to one exponential using the Baker-Campbell-Hausdorff formula. Written out explicitly this is
	\begin{eqnarray}\label{BCH}
	e^{X}e^{Y}=\sum_{n = 1}^\infty\frac {(-1)^{n-1}}{n}
	\sum_{\begin{smallmatrix} r_1 + s_1 > 0 \\ \vdots \\ r_n + s_n > 0 \end{smallmatrix}}
	\frac{[ X^{r_1} Y^{s_1} X^{r_2} Y^{s_2} \dotsm X^{r_n} Y^{s_n} ]}{(\sum_{j = 1}^n (r_j + s_j)) \cdot \prod_{i = 1}^n r_i! s_i!},
	\end{eqnarray}
	where the notation means
	\begin{eqnarray}\label{notation}
	[ X^{r_1} Y^{s_1} \dotsm X^{r_n} Y^{s_n} ] = [ \underbrace{X,[X,\dotsm[X}_{r_1} ,[ \underbrace{Y,[Y,\dotsm[Y}_{s_1} ,\,\dotsm\, [ \underbrace{X,[X,\dotsm[X}_{r_n} ,[ \underbrace{Y,[Y,\dotsm Y}_{s_n} ]]\dotsm]].
	\end{eqnarray}
	To simplify things we redefine the variables $\phi\to \sqrt{K}\phi$ and $\theta \to \theta/\sqrt{K}$ and define 
	\begin{eqnarray}
	X&=&\int\mathrm{d}x\, \left[\nabla \phi(x)\right]^2+\left[\nabla \theta(x)\right]^2\\
	Y&=&\int \mathrm{d}x\,\cos{(qx)}\nabla\phi(x)
	\end{eqnarray}
	Now we will work out all the commutators necessary for the BCH formula. There are only 3 basic ones we need
	\begin{eqnarray}
	[X,Y]&=&\int \mathrm{d}x\mathrm{d}y\cos{(q x)}\,[\left(\nabla \theta(y)\right)^2,\nabla\phi(x)]\\
	&=&2\int \mathrm{d}x\mathrm{d}y\,\cos{(q x)}\left[-i\pi\partial_x\delta(y-x)\right]\nabla\theta(y)\\
	&=&-2i\pi q\int \mathrm{d}x\sin{(qx)}\nabla\theta(y)\\
	&\equiv& -2i\pi q Z
	\end{eqnarray}
	Which defines $Z$. Next we have
	\begin{eqnarray}
	[Y,[X,Y]]&=&-2i\pi q\int \mathrm{d}x\mathrm{d}y\cos{(qx)}\sin{(qy)}[\nabla \phi(x),\nabla\theta(y)]\\
	&=&2\pi^2 q\int \mathrm{d}x\mathrm{d}y\cos{(qx)}\sin{(qy)}\partial_x \delta(y-x)\\
	&=&\pi^2q^2L
	\end{eqnarray}
	where we used the periodic boundary conditions. Lastly we have
	\begin{eqnarray}
	[X,[X,Y]]&=&-2i\pi q\int \mathrm{d}x\mathrm{d}y\sin{(q y)}[(\nabla \phi(x))^2,\nabla\theta(y)]\\
	&=&-(2\pi)^2 q\int \mathrm{d}x\mathrm{d}y\sin{(q y)}\left[\partial_y\delta(y-x)\right]\nabla \phi(x)\\
	&=&(2\pi q)^2Y
	\end{eqnarray}
	So apart from the $Z$ which is different we only generate constants and $X$s.
	Using these we determine that the only non zero commutators appearing in the BCH formula are (in the notation of \eqref{notation})
	\begin{eqnarray}
	[\,X^{n}Y]=\begin{cases}
	-i(2\pi q)^{n}Z~&\text{for}~n ~\text{odd}\\
	~~(2\pi q)^{n}Y~&\text{for}~n ~\text{even}
	\end{cases}
	\end{eqnarray}
	The only other type which is nonzero is
	\begin{eqnarray}
	[YX^{n}Y]=
	(2\pi q)^{n+1}\left(\frac{ L}{4}\right)~&\text{for}~n ~\text{odd}
	\end{eqnarray}
	Everything else vanishes. These three non zero types of commutators can then be inserted into \eqref{BCH} to give the result. Restoring the constants but keeping our redefined bosonic fields we have
	\begin{eqnarray}\label{U}
	H_\text{F}&=&H_\text{Lutt}T+\alpha(T)\int\mathrm{d}x\,\cos{(qx)}\nabla\phi(x)+\beta(T)\int\mathrm{d}x\sin{(q x)}\nabla\theta(x)+\gamma(T)
	\end{eqnarray}
	where the constants introduced above are given by 
	\begin{eqnarray}
	\alpha(T)&=&\sqrt{K}\frac{V}{\pi}\frac{\sin{(v_s q T )}}{v_sq T }\\
	\beta(T)&=&\sqrt{K}\frac{V}{\pi}\frac{1-\cos{(v_sqT)}}{v_sq T}\\
	\gamma(T)&=&\left(\frac{\sqrt{K}V}{v_sT}\right)^2\frac{L}{2\pi q}\left[\sin{(v_sqT)}-v_sqT\cos{(v_sqT)}\right]
	\end{eqnarray}
	It is easy to check that upon taking either $T\to 0$ or $V \to 0$ we recover the expected result. 
	
	%%%%%%%%%%%%
	\subsection{Energy at stroboscopic times }
	We now calculate energy after the $N^\text{th}$ kick, $t=NT$. This is given by 
	\begin{eqnarray}\label{strobeenergy}
	E(t)
	&=&\matrixel{\Psi_0}{e^{iNH_\text{F}}He^{-iNH_\text{F}}}{\Psi_0}
	\end{eqnarray}
	To proceed we perform the following shift of the field operators
	\begin{eqnarray}
	\tilde{\phi}(x)&=&\phi(x)-\frac{\pi\alpha(T)}{q v_s T}\sin{(qx)}\\
	\tilde{\theta}(x)&=&\theta(x)+\frac{\pi\alpha(T)}{qv_sT}\cos{(qx)}\\
	H_\text{F}&=&\frac{v_sT}{2\pi}\int\mathrm{d}x [\nabla\tilde{\phi}(x)]^2+[\nabla\tilde{\theta}(x)]^2-\frac{\pi L}{4v_s T}(\alpha^2(T)+\beta^2(T))
	\end{eqnarray}
	using this in \eqref{strobeenergy} we have that the energy is
	\begin{eqnarray}\label{energy1}
	E(t)&=&\matrixel{\Psi_0}{H_\text{F}/T}{\Psi_0}+\frac{\pi L}{4uT^2 }(\alpha^2(T)+\beta^2(T))\\
	&&+\frac{\alpha(T)}{T}\int \mathrm{d}x \sin{(qx)}\matrixel{\Psi_0}{e^{iNH_\text{F}}\nabla\tilde{\theta}(x)e^{-iN H_\text{F}}}{\Psi_0}\\
	&&+\frac{\beta(T)}{T}\int \mathrm{d}x \cos{(qx)}\matrixel{\Psi_0}{e^{iN H_\text{F}}\nabla\tilde{\phi}(x)e^{-iN H_\text{F}}}{\Psi_0}
	\end{eqnarray}
	Now it is convenient to introduce the mode expansions for the field variables,
	\begin{eqnarray}
	\nabla\tilde{\phi}(x)&=&-\frac{\pi N}{\sqrt{K}L}-\frac{1}{2}\sum_{p\neq 0}\left(\frac{2\pi |p|}{L}\right)^\frac{1}{2}e^{-ipx}\left[b_{p}^\dag+b_{-p}\right]\\
	\nabla\tilde{\theta}(x)&=&\frac{\pi \sqrt{K} J}{L}+\frac{1}{2}\sum_{p\neq 0}\left(\frac{2\pi |p|}{L}\right)^\frac{1}{2}e^{-ipx}\text{sgn}(p)\left[b_{p}^\dag-b_{-p}\right]\\
	H_\text{F}/T&=&\sum_{p\neq 0}v_s|p|b^\dag_pb_p+\frac{\pi v_s}{2KL}\mathcal{N}^2+\frac{K\pi v_s}{2L}J^2
	\end{eqnarray}
	where $\mathcal{N}$ is the total particle number and $J$ the total current. Inserting these into \eqref{energy1} and using $e^{iH_\text{eff}}b_p^\dag e^{-iH_\text{F}}=e^{iv_s|p|T}b^\dag_p$ we get 
	\begin{eqnarray}
	E(t)&=&-\frac{\pi\alpha(T)}{2T}\left(\frac{L |q|}{2\pi}\right)^\frac{1}{2}\left[e^{iNv_sqT}\left<b^\dag_{q}+b^\dag_{-q}\right>+e^{-iNv_sqT}\left<b_{q}+b_{-q}\right>\right]
	\\
	&&-i\frac{\pi\beta(T)}{2T}\left(\frac{L |q|}{2\pi}\right)^\frac{1}{2}\left[e^{iNv_sqT}\left<b^\dag_{q}+b^\dag_{-q}\right>-e^{-iNv_sqT}\left<b_{q}+b_{-q}\right>\right]\\
	&&+\matrixel{\Psi_0}{H_\text{F}(T)}{\Psi_0}+\frac{\pi L}{4v_sT^2 }(\alpha^2(T)+\beta^2(T))
	\end{eqnarray}
	These expectations values do not vanish as $\ket{\Psi_0}$ is a coherent state when written in terms of $ b_p^\dag, ~b_p$. To see this we write $H$ in this fashion,
	\begin{eqnarray}
	H&=&H_\text{F}/T+\frac{\alpha(T)}{T}\int\cos{(qx)}\nabla\tilde{\phi}(x))+\frac{\beta(T)}{T}\int\sin{(qx)}\nabla\tilde{\theta}(x)+\frac{\pi L}{4u T^2}\left[\alpha(T)^2+\beta(T)^2\right]\\
	&=&H_\text{F}/T-X(q,T)\left[b^\dag_q+b^\dag_{-q}\right]-X(q,T)^*\left[b_q+b_{-q}\right]
	\end{eqnarray}
	where 
	\begin{eqnarray}
	X(q,T)=\frac{\pi}{2T}\left(\frac{L|q|}{2\pi}\right)^{\frac{1}{2}}\left[\alpha(T)+i\beta(T)\right]
	\end{eqnarray}
	In terms of the $b_p$s  the ground state of $H$ is therefore
	\begin{eqnarray}
	\ket{\Psi_0}=Ce^{-\frac{X(q,T)}{v_s|q|}\left[b_q^\dag+b^\dag_{-q}\right]}\ket{0}
	\end{eqnarray}
	with $C=\exp{\{-|X|^2/(2v_s^2|q|^2)\}}$ the normalisation. The expectation values of the operators are in the initial state are 
	\begin{eqnarray}
	\left<b_{q}+b_{-q}\right>=2X(q,T),~~\left<b^\dag_{q}+b^\dag_{-q}\right>=2X^*(q,T)
	\end{eqnarray}
	Combining all this together we get that the stroboscopic energy is 
	\begin{eqnarray}\label{ENERGY}
	E(t)
	&=&\frac{KV^2L}{v_s\pi T^2}\left[\frac{\sin^2{\left(\frac{v_sqT}{2}\right)}}{v_sqT}\right]\left[\frac{\sin^2{\left(N\frac{v_sqT}{2}\right)}}{v_sqT}\right]
	\end{eqnarray}
	where $t=NT$.
	
	\subsection{Density and Current}
	Within the same formalism we can calculate the expectation value of the current and density which apart form being of interest are necessary self consistency checks. Using the various manipulations above we find that at stroboscopic times expectation values of the density, $\rho(x,N)$ and current, $J(x,N)$ are  given by 
	\begin{eqnarray}\label{density}
	\left<\rho(x,N)\right>&=&\rho_0+\frac{\pi \alpha(T)}{v_sT}\cos{(qx)}-\frac{\pi}{v_sT}\cos{(qx)}\left[\alpha(T)\cos{(v_sNTq)}+\beta(T)\sin{(v_sNTq)}\right]\\\label{current}
	\left<J(x,N)\right>&=&\frac{\pi \beta(T)}{v_sT}\sin{(qx)}-\frac{\pi}{v_sT}\sin{(qx)}\left[\alpha(T)\cos{(v_sNTq)}-\beta(T)\sin{(v_sNTq)}\right]
	\end{eqnarray}
	both of these expression recover the expected results in the trivial limits ($T,N, V\to 0$) and also are consistent within our model as the density only varies slowly over the system. Taking a long time average, $\lim_{t\to\infty}1/t\int^t \mathrm{d}t'$  we see that the system displays periodic density and current variations
	\begin{eqnarray}
	\left<\left<\rho(x,NT)\right>\right>&=&\rho_0+\frac{\pi \alpha(T)}{v_sT}\cos{(qx)}\\
	\left<\left<J(x,NT)\right>\right>&=&\frac{\pi \beta(T)}{v_sT}\sin{(qx)}
	\end{eqnarray}
	
	%%%%%%%%%%%%%%
	
	%%%%%%%%%%%%%%
	\subsection{Wigner Function approach}
	Within the fermionic formulation of the model we can examine the system using the Wigner function approach as we did for the TG gas above. Here we define a Wigner function for each of the branches, $\sigma=\pm$, as 
	\begin{eqnarray}
	n_\sigma(x,\lambda)=\int\mathrm{d}\lambda\,e^{i\lambda y}\left<\check{\psi}_\sigma^\dag(x+y/2)\check{\psi}_\sigma(x-y/2)\right>
	\end{eqnarray}
	and as before upon integrating over $x$ or $\lambda$ this gives the momentum distribution function or spatial density respectively.
	Between kicks this satisfies the equation of motion
	\begin{eqnarray}
	\left[\partial_t+\sigma v_s\partial_x\right]n_\sigma(x,\lambda,t)=0.
	\end{eqnarray}
	which is solved by $n_\sigma(c,\lambda,t+T^-)=n_\sigma(x-v_sT,\lambda,t)$. Through the kicks we can use $e^{-iH_\text{K}}\check{\psi}_\sigma^\dag(x)e^{iH_\text{K}}=e^{-iV\cos{qx}}\check{\psi}_\sigma^\dag(x)$ to get that
	\begin{eqnarray}
	\tilde{n}_\sigma(x,z,t+T^+)=e^{2iV\sin{(qz/2)}\sin{(qx)}}\tilde{n}_\sigma(x,z,t+T^-)
	\end{eqnarray}
	where $\tilde{n}_\sigma$ denotes the Fourier transform with respect to $\lambda$. 
	This is the same evolution as the TG case but with $v_\text{eff}=\sigma v_s$. However in this instance as $v_\text{eff}$ is independent of the momentum it can be determined analytically.
	Staring from  distribution that is independent of $x$, $n_\pm(x,\lambda,0)=n_0(\lambda)$ we have that 
	\begin{eqnarray}
	n_\sigma(x,\lambda,NT)=\sum_{n_1\dots n_N=-\infty}^\infty\prod_{l=1}^NJ_{n_l}\left(2V\sin{\big\{q(x-\sigma(N-l+1)v_sT)\big\}}\right) n_0\Big(\lambda+\sum_{j=1}^Nn_jq/2\Big)
	\end{eqnarray}
	where $J_n(x)$ is a Bessel function.
	The total energy is given by $\int\mathrm{d}x\mathrm{d}\lambda\, v_s\lambda \left[n_+(x,\lambda,t)+n_-(x,\lambda,t)\right]$. 
	%%%%%%%%%%%%%%%
	
	%%%%%%%%%%%%%%%%
% 	\section{Non Linear Luttinger Liquid}
% 	The non linear theory is described by the Hamiltonian
% 	\begin{eqnarray}\label{HnLLS}
% 	H_\text{nL}=\sum_{\sigma=\pm}\int\mathrm{d}x\,:\psi_{\sigma}^\dag(x)\left[-i\sigma v_s\partial_x-\frac{\partial^2_x}{2m^*}\right]\psi_\sigma(x):+\sum_{j=0}^\infty\delta(t-jT)\sqrt{K}V\cos{(qx)}\rho(x).
% 	\end{eqnarray}
% 	which describes the system beyond the linear regime. We can investigate this by follow the evolution of the Wigner function of this model as we did before. Once again we define a Wigner function for each branch and find that
% 	\begin{eqnarray}
% 	\left[\partial_t+v_\text{eff}(\lambda)\partial_x\right]n_\sigma(x,\lambda,t)=0
% 	\end{eqnarray}
% 	where now $v_\text{eff}(\lambda)=\sigma v_s+\lambda/m^*$ while through the kick the we have the same behavior. The energy of the system is given by 
% 	$$\int\mathrm{d}x\mathrm{d}\lambda\, \left(v_s\lambda+\lambda^2/2m^*\right) \left[n_+(x,\lambda,t)+n_-(x,\lambda,t)\right].$$
% 	The linear theory is reproduced by taking $m^*\to\infty$ while if instead we take $v_s\to0$ we see that the system behaves as two copies of the TG gas. Therefore the non linear theory also exhibits dynamical localization. 

	\begin{figure}
		\centering
		\includegraphics[width=0.5\linewidth]{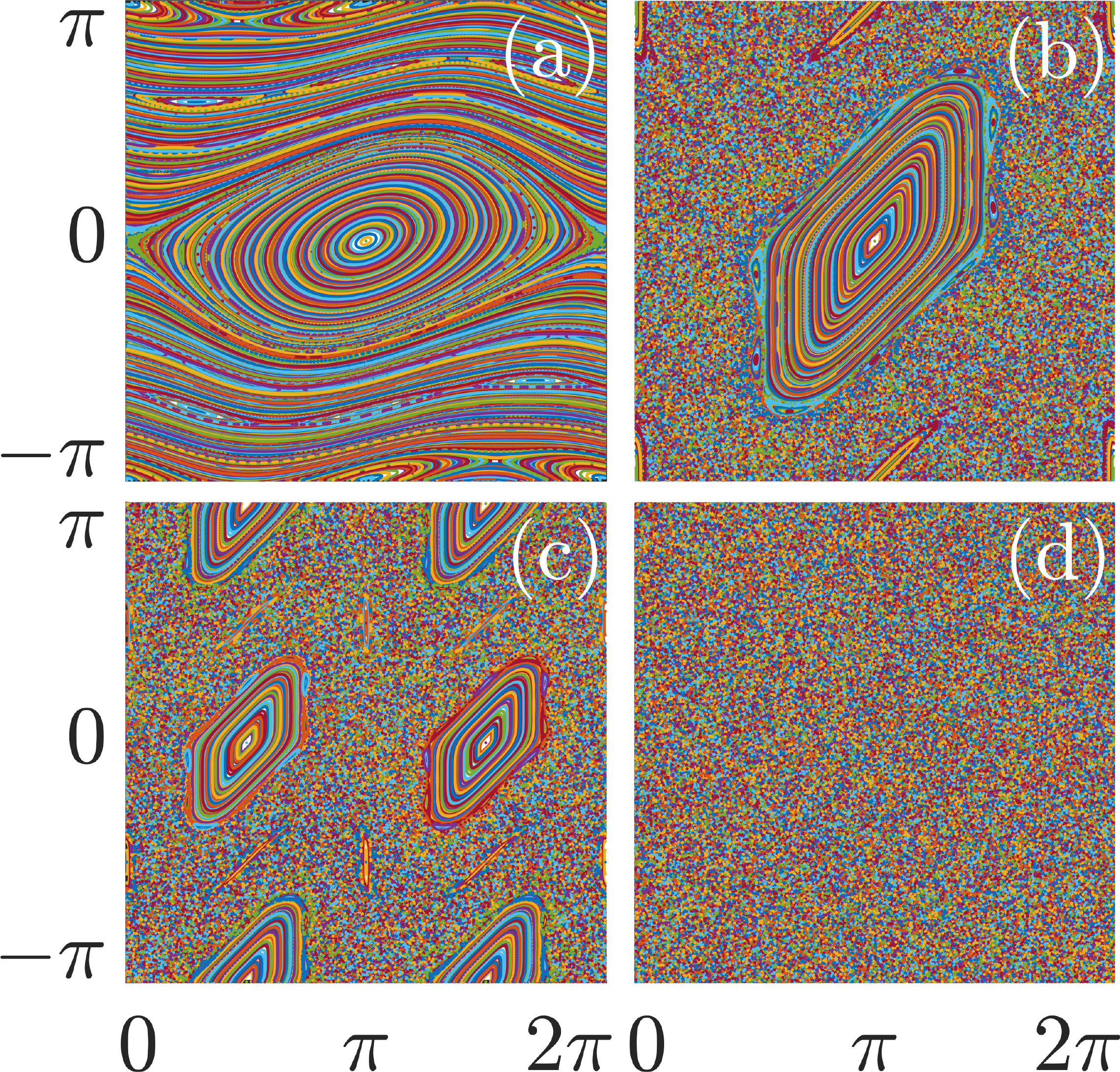}
		\caption{The phase portrait of Chirikov's standard map. (a) $V=0.5$, $q=1$; (b) $V=2$, $q=1$; (c) $V=0.5$, $q=2$; (d) $V=2$, $q=2$. { The parameters as in panel (c) are used in the main text for the kicked Lieb-Liniger model.}}
		\label{fig:KR_PhasePort}
	\end{figure}

\section{3. Classical Kicked Rotor under a non-$2\pi$-periodic kicking potential}
	
In this section, we show that the effective kicking strength in the classical single-particle analog of our model -- the Chirikov's standard map -- is modified. In particular, the Hamilton's equations of motion read:
	
\begin{equation}
	\left\{
	\begin{array}{ll}
	\vspace{5pt}  p_{n+1}=p_n+qV\sin(qx_n), \mod 2\pi\\
	x_{n+1}=x_n+p_{n+1}, \mod 2\pi
	\end{array}
	\right.,
\end{equation}
where we adopt the units in which $x_n\in[0, 2\pi)$, $q\in\mathbb{Z}$, and, as one can check, both $x_n$ and $p_n$ are $2\pi/q$-periodic. Conventionally, $q=1$. In case of general $q$, though, one can make a coordinate change: $\tilde{p}_n=qp_n$, $\tilde{x}_n=qx_n$ with the new coordinates $\tilde{x}_n \in [0, 2\pi q)$, which are $2\pi$-periodic. In these coordinates, the equations read:
\begin{equation}
	\left\{
	\begin{array}{ll}
	\vspace{5pt}  \tilde{p}_{n+1}=\tilde{p}_n+q^2V\sin(\tilde{x}_n), \mod 2\pi q\\
	\tilde{x}_{n+1}=\tilde{x}_n+\tilde{p}_{n+1}, \mod 2\pi q
	\end{array}
	\right.,
\end{equation}
and it is customary to reduce the consideration to region of periodicity $\tilde{x}_n, \tilde{p}_n \in [0, 2\pi)$. In these coordinates, the conventional standard map at $q=1$ is restored with the kicking strength parameter $\tilde{V} = q^2V$.  For example, at $q=2$, kicking at $V=0.5$ is equivalent to kicking the conventional map at $q=1$ with $\tilde{V}=4V=2$ -- well above the regular-to-chaotic transition at $V_{\rm cr}\approx0.97$. Fig.~\ref{fig:KR_PhasePort} demonstrates this correspondence.

\section{4. Additional data and approximation justification}

\begin{figure}
    \centering
    \includegraphics[width=0.7\linewidth]{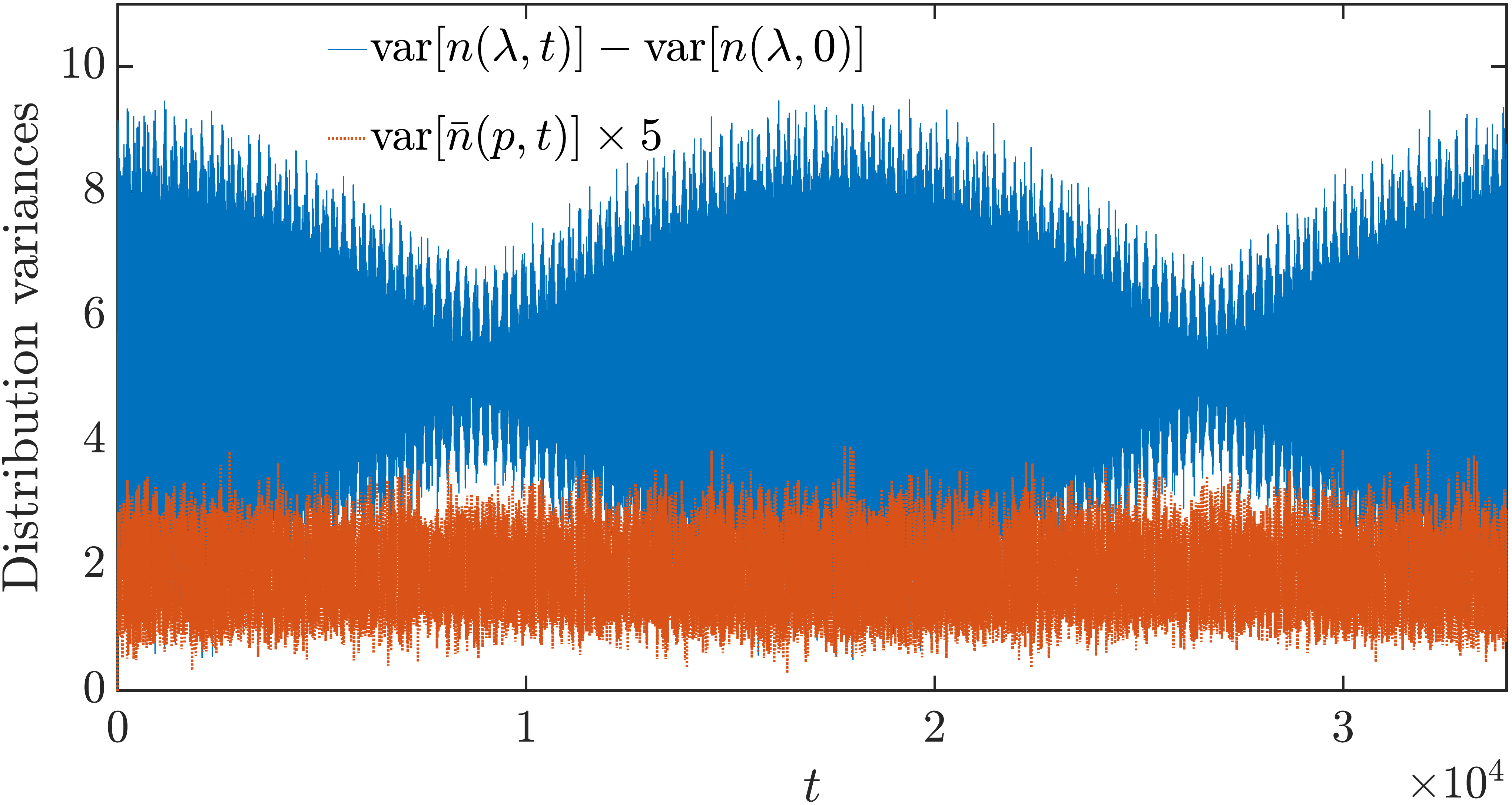}
    \caption{Upper solid blue curve: variance of the momentum density $n(\lambda,t)$ in the kicked Lieb-Liniger gas as a function of time relative to the initial variance: ${\rm var}[n(\lambda,t)]-{\rm var}[n(\lambda,0)]$. It has oscillatory character that hints on certain underlying invariance in the system (possibly approximate) and  hence at least transient dynamical localization. Lower dotted red curve: scaled variance of $\bar{n}(p,t)$. In this parameter regime, we reach its saturation, and GHD and our scheme may be applicable for long times without eventual delocalization.
	Parameters: $V=0.15,\; q=4\pi/L,\; \gamma=10,\; \mathcal{N}=200$.}
    \label{fig:LL_Energy_dPsat}
\end{figure}

\begin{figure}
    \centering
    \includegraphics[width=0.7\linewidth]{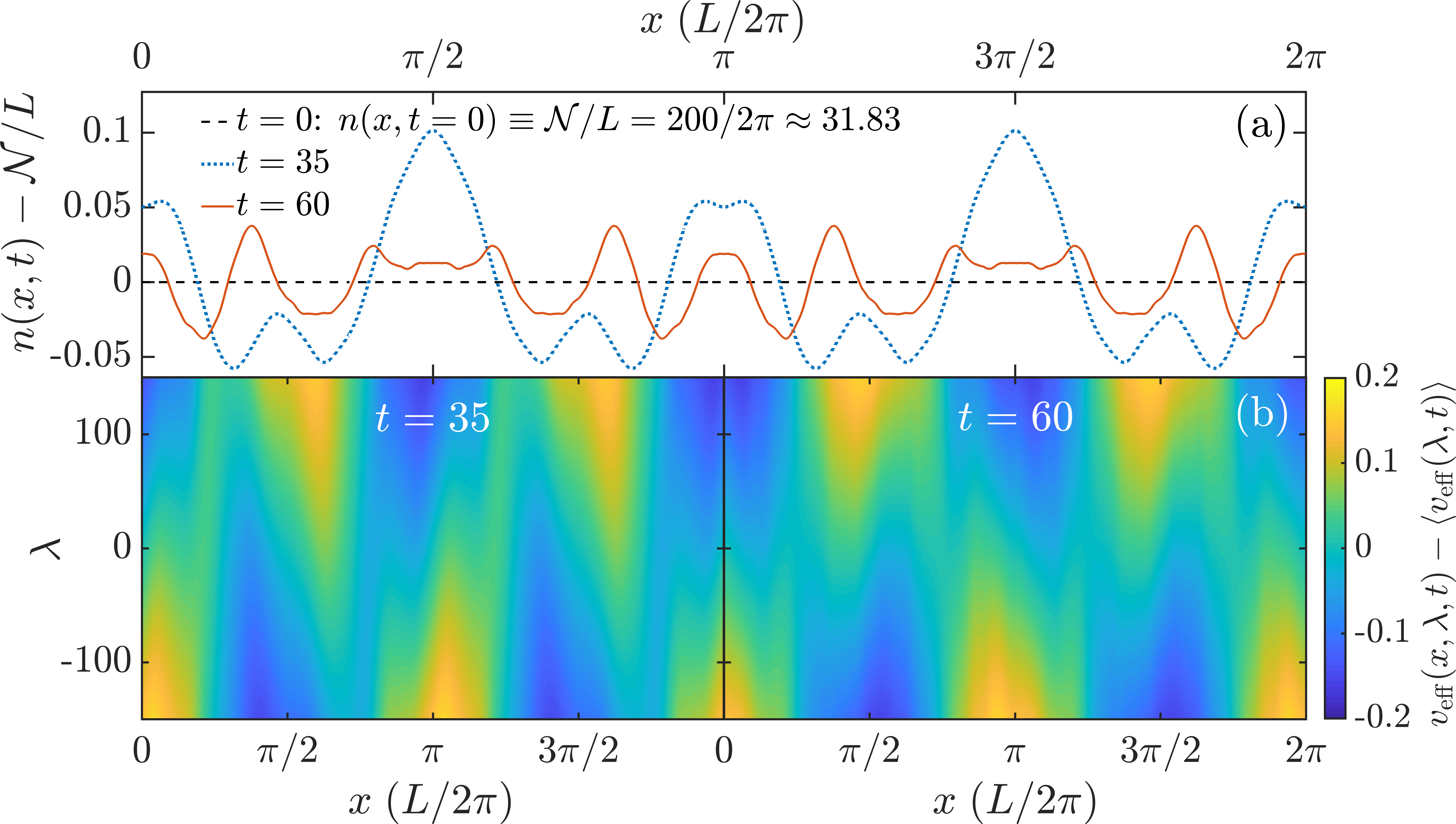}
    \caption{(a) Spatial density profile, $n(x,t) = \int \frac{d\lambda}{2\pi} n(x,\lambda,t)$, calculated at two different times using the exact scheme from the main text. Both cases demonstrate less than $0.4\%$ deviation from the constant average density of $\mathcal{N}/L = 200/2\pi \approx 31.83$.\\ (b) The deviation of the effective velocity from its mean value, $\veff(x,\lambda,t)-\langle\veff(\lambda,t)\rangle$, where $\langle\ldots\rangle$ represents averaging over the position $x$. Obtained via the exact method discussed in the main text. Left and right panels show examples at the same times as in panel (a).
    Parameters are the same as in the main text: $V=0.5$, $q=4\pi/L$, $\gamma=10$.}
    \label{fig:n_veff_x}
\end{figure}

In this section, we have two goals. First, in Fig.~\ref{fig:LL_Energy_dPsat}, we demonstrate the regime, in which the variance of the Fourier transform of the spatial density is saturated, which is achieved at weaker kicking. Second, we show that our approximate scheme is justified. For that purpose, we employ the exact scheme described in the main text in order to compute the spatial density and effective velocity at short times and demonstrate the they are close to being constants with respect to the corresponding scales. In particular, in Fig.~\ref{fig:n_veff_x}(a), the spatial density is shown to vary across the system by less than $0.4\%$ of its average value.
% In Fig.~\ref{fig:n_veff_x}(b), the corresponding effective velocity is shown to vary within $0.5\%$ of its average value, which is of the same order as $\lambda$ (vertical axis).

\end{document}